\documentclass[aps,twocolumn,amsmath,superscriptaddress,longbibliography]{revtex4-1}
\usepackage{graphicx,epstopdf,bm}
\usepackage{physics}

\makeatletter
\def\@bibdataout@aps{%
	\immediate\write\@bibdataout{%
		@CONTROL{%
			apsrev41Control%
			\longbibliography@sw{%
				,author="08",editor="1",pages="1",title="0",year="1"%
			}{%
				,author="08",editor="1",pages="1",title="",year="1"%
			}%
		}%
	}%
	\if@filesw \immediate \write \@auxout {\string \citation {apsrev41Control}}\fi 
}
\makeatother

\newcommand{\nobracket}{}

\newcommand{\tmmathbf}[1]{\ensuremath{\boldsymbol{#1}}}
\newcommand{\tmop}[1]{\ensuremath{\operatorname{#1}}}

\newcommand{\mbf}{\boldsymbol}
\newcommand{\vep}{\varepsilon}
\newcommand{\rNum}[1]{\lowercase\expandafter{\romannumeral #1\relax}}
\newcommand{\RNum}[1]{\uppercase\expandafter{\romannumeral #1\relax}}

\newcommand{\Rmnum}[1]{\expandafter\@slowromancap\romannumeral #1@}

\usepackage[usenames,dvipsnames,svgnames,table]{xcolor}
\usepackage[colorlinks=true,linkcolor=RoyalBlue,citecolor=RoyalBlue]{hyperref}

\begin{document}

\title{Electric Polarization in Inhomogeneous Crystals}

\author{Yiqiang Zhao}
\affiliation{International Center for Quantum Materials, School of Physics, Peking University, Beijing 100871, China}
\affiliation{Department of Physics, Carnegie Mellon University, Pittsburgh, Pennsylvania 15213, USA}

\author{Yang Gao}
\affiliation{Department of Physics, Carnegie Mellon University, Pittsburgh, Pennsylvania 15213, USA}

\author{Di Xiao}
\affiliation{Department of Physics, Carnegie Mellon University, Pittsburgh, Pennsylvania 15213, USA}

\date{\today}

\begin{abstract}
	We derive the charge density up to second order in spatial gradient in inhomogeneous crystals using the semiclassical coarse graining procedure based on the wave packet method. It can be recast as divergence of polarization, whose first-order contribution consists of three parts, a perturbative correction to the original Berry connection expression, a topological part that can be written as an integral of the Chern-Simons 3-form, and a previously-unknown, quadrupole-like contribution. The topological part can be related to the quantized fractional charge carried by a vortex in two-dimensional systems. We then generalize our results to the multi-band case and show that the quadrupole-like contribution plays an important role, as it makes the total polarization gauge-independent. Finally, we verify our theory in several model systems.

\end{abstract}

{\maketitle}

\section{Introduction}

The electric polarization $\bm P$ is an essential quantity in the macroscopic theory of electromagnetism.  Its spatial and temporal dependence give rise to the implicit charge density $\rho$ and current density $\bm j$ carried by the medium via the following relations,
\begin{subequations} \label{P_def}
\begin{align}
\rho &= -\bm\nabla \cdot \bm P \;, \label{P_density} \\
\bm j &= (\partial/\partial t)\bm P \;. \label{P_current}
\end{align}
\end{subequations}
These relations can be rigorously established through a spatial averaging procedure known as coarse graining, which is designed to produce spatially slowly varying macroscopic quantities from their rapidly varying microscopic counterparts (see, for example, Sec.~6.6 of Ref.~\cite{Jackson1999}).  However, despite its apparent simplicity in appearance, calculating $\bm P$ for a given microscopic charge density of an extended system has proven to be problematic.  In fact, it has been shown that one cannot calculate the polarization from the microscopic charge density alone.  Instead, Eq.~\eqref{P_def} should be imposed as the fundamental definition of $\bm P$ and consequently the starting point of any microscopic theory.  In the modern theory of electric polarization~\cite{King-Smith1993,vanderbilt1993,Resta1994}, Eq.~\eqref{P_current} is used to relate $\bm P$ to the integral of the adiabatic current~\cite{Thouless1983}. The resulting expression of $\bm P$ is given in terms of the Berry connection of the Bloch functions~\cite{King-Smith1993,vanderbilt1993,Resta1994}.  This theory has been very successful in understanding dielectric phenomena, and also forms an essential part in our understanding of topological materials. 

The modern theory of electric polarization is developed for perfect crystals, i.e., crystals with translational symmetry.  The purpose of this paper is to develop a general theory of electric polarization in inhomogeneous crystals.  Here, by inhomogeneous crystals we mean crystals under the influence of external perturbations that break translational symmetry and vary slowly in space.  Our motivation is two-fold.  First, as inhomogeneity frequently occurs in condensed matter systems, this problem appears in a wide range of physical applications.  There have already been quite a few studies of polarization induced by particular types of inhomogeneities, such as strain~\cite{Martin1972,Nelson1976,DalCorso1994,Bernardini1997,Saghi-Szabo1998,Vanderbilt2000,Bellaiche2000,Liu2017b,Hong2013}, strain gradient~\cite{Resta2010,Hong2011,Hong2013,Stengel2013,Schiaffino2019}, electromagnetic fields~\cite{Essin2009,Essin2010,Gao2014,Coh2011,Bousquet2011,Malashevich2012,Mostovoy2010,Malashevich2010}, and spin textures in multiferroics~\cite{Lawes2005,Kenzelmann2005,Neaton2005,Katsura2005,Jia2006,Mostovoy2006,Jia2007,Harris2007,Kenzelmann2007,Malashevich2008,Malashevich2009,Xiang2011,Xiang2013}. However, despite an early attempt~\cite{Xiao2009}, a complete and unified theory appropriate for any type of spatial inhomogeneity is still absent.  Second, the interpretation of the polarization in terms of the adiabatic current [Eq.~\eqref{P_current}] has been the dominant approach in formal theory development. Here we introduce an alternative approach to calculate $\bm P$ using Eq.~\eqref{P_density} as the starting point. We note that the polarization from Eq.~\eqref{P_density} is equivalent to that from Eq.~\eqref{P_current}, due to the continuity equaiton:
\begin{equation}
(\partial  /\partial t)\rho +\nabla \cdot \bm j=0.
\end{equation}	

The key to our approach is a semiclassical coarse graining procedure based on the framework of wave packet dynamics of Bloch electrons~\cite{Sundaram1999,Xiao2010}, which allows us to directly calculate the charge density $\rho(\bm r)$ in an order-by-order fashion.  We can then extract the polarization from $\rho(\bm r)$ according to Eq.~\eqref{P_density}.  In this alternative approach, the polarization charge density becomes the central quantity, which avoids many conceptual difficulties.

With the semiclassical coarse graining procedure, we derive the charge density up to second order in spatial gradient, which requires us to first generalize the semiclassical theory of electron dynamics to second order.  Note that there are both ionic and electronic contributions to the charge density and we are only concerned with the latter.  We show that the charge density can be reformulated using the electric polarization up to first order. At zeroth order, the polarization from Eq.~\eqref{P_density} indeed coincides with that from Eq.~\eqref{P_current}, confirming the relationship between polarization and charge density in extended systems. At first order, the polarization consists of three parts, a perturbative correction to the original Berry connection expression, a topological part that can be written as an integral of the Chern-Simons 3-form, and a previously-unknown, quadrupole-like contribution. We show that in two-dimensional systems, the topological part can be related to the fractional charge carried by a vortex. We also generalize our results to the multi-band case, in which we find that the quadrupole-like contribution is indispensable as it makes the total polarization gauge-independent.

To further establish the validity and utility of our theory, we apply it to several examples.  We first consider an exactly solvable problem, i.e., the change of charge density due to a constant strain, and show that our theory is consistent with the exact result up to second order.  We then numerically test our theory in a one-dimensional modified Su-Schrieffer-Heegar (SSH) model and a two-dimensional $\pi$-flux model on a square lattice.  In the 1D model, we verify the non-topological contribution of first-order polarization and discuss how the coarse graining procedure should be carried out in the numerical simulation.  In the 2D model, we verify the topological contribution in our theory, and relate it to the appearance of quantized fractional charge carried by a vortex.

Our paper is organized as follows. We present our formalism in Sec.~\ref{Gen}, which contains a detailed application of the coarse graining method, its applications in Sec.~\ref{Num}, and conclude with a summary in Sec.~\ref{Summary}.

\section{General formulation}

\label{Gen}

The theory of electric polarization was previously developed using the concept of adiabatic current~\cite{King-Smith1993,Resta1994}.  Here we take a different route and derive it from the charge density using Eq.~\eqref{P_density}.  Specifically, we will calculate the charge density up to second order in spatial gradient, which then allows us to extract the electric polarization up to first order.  For this purpose, we extend the semiclassical theory of wave packet dynamics to second order in Sec.~\ref{GenA}.  We then use it to derive the charge density via the coarse graining procedure in Sec.~\ref{GenB}.  In Sec.~\ref{GenC}, we show that this charge density can be readily recast using the electric polarization, whose first-order term consists of three contributions: a perturbative, a topological, and a quadrupole-like contribution. Finally, we extend our results to the multi-band case in Sec.~\ref{GenD}.

\subsection{Semiclassical theory up to second order}

\label{GenA}

To set up the notation, we first briefly review the semiclassical theory of wave packet dynamics.  For details we refer the readers to Ref.~\cite{Sundaram1999,Xiao2010}. Let us consider an insulating crystal with slowly varying inhomogeneities described by the Hamiltonian $\hat{H}(\hat{\bm r}, \hat{\bm p};\beta_i (\hat{\bm r}))$, where $\beta_i (\hat{\bm r})$ are a set of slowly varying parameters characterizing the inhomogeneities. They may represent strain fields, electromagnetic fields, spin textures, and so on.  The exact Hamiltonian is difficult to diagonalize because the translational symmetry is broken by $\beta_i(\hat{\bm r})$.  Instead, we can simplify this problem by taking a wave packet localized around $\bm r_c$ as an approximate solution. We assume that the spread of the wave packet is small compared to the length scale of the spatial inhomogeneity such that its dynamics is governed by a
local Hamiltonian $\hat{H}_c (\tmmathbf{r}_c) = \hat{H} (\hat{\tmmathbf{r}}, \hat{\tmmathbf{p}};
\beta_i(\tmmathbf{r}_c))$ at the leading order. The local Hamiltonian $\hat{H}_c$  is obtained by replacing $\beta_i({\bm{\hat{r}}})$ with their value at $\bm{r}_c$ in exact Hamiltonian $\hat{H}$. In this way, the translational symmetry is restored. Throughout this paper, order in spatial gradient means order in $\partial_{r_{ci}}\beta_j(\bm{r}_c)$, which is a small quantity by our assumption. Higher order contributions can be obtained by including higher order terms in the expansion of the Hamiltonian $\hat{H}(\hat{\bm r}, \hat{\bm p};\beta_i (\hat{\bm r}))$ around $\bm r_c$ in $\beta_i(\hat{\bm r})$.

In the following we shall focus on a single, non-degenerate band with band index $0$.  The wave packet $\ket{W (\tmmathbf{r}_c, \tmmathbf{k}_c)}$ is constructed from the local Bloch function $\ket{\psi_{0\boldsymbol{k}}(\mbf{r}_c)} = e^{i\mbf{k}\cdot\mbf{r}}\ket{u_{0\mbf{k}}(\mbf{r}_c)}$,
\begin{equation}
\ket{W (\tmmathbf{r}_c, \tmmathbf{k}_c)}=\int d\mbf{k}\,C_0(\mbf{k}) e^{i\mbf{k}\cdot\mbf{r}}\ket{u_{0\mbf{k}}(\mbf{r}_c)},
\end{equation}
where the expansion coefficient $C_0(\bm k)$ is sharply centered around $\bm k_c$, with its phase fixed through the self-consistency condition: $\bra{W}\hat{\mbf{r}}\ket{W}=\mbf{r}_c$.  In actual calculations, we can approximate $|C_0(\mbf{k})|^2\approx \delta(\mbf{k}-\mbf{k}_c)$. 

Using the time-dependent variational principle, one can work out the equations of motion for $\mbf{r}_c$ and $\mbf{k}_c$~\cite{Sundaram1999},
\begin{subequations} \label{eq_EOM}
\begin{align}
\dot{r}_{ci}&=\frac{\partial \tilde{\vep}_0}{\partial k_{ci}}-\Omega_{k_{ci}r_{cj}}\dot{r}_{cj}-\Omega_{k_{ci}k_{cj}}\dot{k}_{cj},\label{eq_EOM1}\\
\dot{k}_{ci}&=-\frac{\partial \tilde{\vep}_0}{\partial r_{ci}}+\Omega_{r_{ci}r_{cj}}\dot{r}_{cj}+\Omega_{r_{ci}k_{cj}}\dot{k}_{cj},\label{eq_EOM2}
\end{align}
\end{subequations}
where $\tilde{\vep}_0$ is energy of the wave packet, which is the expectation value of exact Hamiltonian $\hat{H}$ on the wave packet, and we have set $\hbar=1$.  The Berry curvature $\Omega_{\xi_i \xi_j}$ is defined by
\begin{equation}
\Omega_{\xi_i \xi_j}=i\bra{\partial_{\xi_i}u_0}\ket{\partial_{\xi_j}u_0}-i\bra{\partial_{\xi_j}u_0}\ket{\partial_{\xi_i}u_0},\label{BC}
\end{equation}
where $|u_0\rangle$ is a shorthand for $|u_{0\bm k_c}(\bm r_c)\rangle$, and $\bm \xi_i = (\bm k_c, \bm r_c)$. Throughout this paper, summation over spatial indices $(i,j,l,t)$ is implied by repeated indices, while summation over band indices $(n,n',m,m')$ is explicitly written.

The appearance of the Berry curvature in the equation of motion Eq.~\eqref{eq_EOM} also has a profound effect on the density of states in the phase space.  Specifically, $\bm r_c$ and $\bm k_c$ are no longer canonically conjugate.  Therefore, one has to introduce a $\bm r_c$- and $\bm k_c$-dependent phase space measure $D(\mbf{r}_c,\mbf{k}_c)$ when taking thermodynamic average in the phase space~\cite{Xiao2005}
\begin{equation}
\iint\frac{d\bm r_c d\bm k_c}{(2\pi)^d}\rightarrow\iint\frac{d\bm r_c d\bm k_c}{(2\pi)^d} D(\bm r_c,\bm k_c) \;,
\end{equation}
where $d$ is the dimension of the system.  The phase space measure, also called the modified density of states, is given by
\begin{align}
D(\mbf{r}_c,\mbf{k}_c) & =  \sqrt{\det (\Omega - J)}, \label{mDos} \\
\Omega - J & =  \left(\begin{array}{cc}
\Omega_{\tmmathbf{r}_c\tmmathbf{r}_c} & \Omega_{\tmmathbf{r}_c\tmmathbf{k}_c} -
I\\
\Omega_{\tmmathbf{k}_c\tmmathbf{r}_c} + I &
\Omega_{\tmmathbf{k}_c\tmmathbf{k}_c}
\end{array}\right),\label{eq_mDos2}
\end{align}
where each block is a $d\times d$ matrix, $I$ is the rank-$d$ identity matrix, and the Berry curvature matrix $\Omega_{\tmmathbf{r}_c\tmmathbf{r}_c}$, $\Omega_{\tmmathbf{r}_c\tmmathbf{k}_c}$, $\Omega_{\tmmathbf{k}_c\tmmathbf{r}_c}$, $\Omega_{\tmmathbf{k}_c\tmmathbf{k}_c}$ are defined above in Eq.~(\ref{BC}).

The above semiclassical theory was originally derived up to first order in spatial gradient. For our purpose, we need to generalize it to second order.  This has been done in Ref.~\cite{Gao2014} for the special case of constant electromagnetic fields.  Following the same procedure outlined in Ref.~\cite{Gao2014}, we find that for a general perturbation, the form of the equation of motion Eq.~\eqref{eq_EOM} remains unchanged.  This implies that the form of the modified density of states in Eq.~\eqref{mDos} is also unchanged.  The modification enters in two places: (i) the energy of the wave packet needs to be modified to include second-order terms.  This modification is irrelevant to our calculation due to the fact that energy correction leads to Fermi surface effect which is zero in insulators and will not be discussed further. (ii) the Berry curvature should be calculated using the periodic part of the perturbed Bloch function $|\tilde{u}_0\rangle$ up to first order in spatial gradient.  Since terms involving the Berry curvature in Eq.~\eqref{eq_EOM} already have at least one explicit spatial derivatives, Bloch functions corrected up to first order are sufficient for a second-order theory.

The exact form of $\ket{\tilde u_0}$ can be determined as follows.  Let $\ket{\tilde{u}_0}=\ket{u_0}+\ket{\delta u_0}$, where $\ket{\delta u_0}$ is the correction to the wave function caused by the first-order correction $\hat{H}'$ to the local Hamiltonian, where $\hat{H}'$ is obtained by the gradient expansion of $\hat H_c$,
\begin{equation}
\hat{H}' = \frac{1}{2} \left[ (\hat{\tmmathbf{r}} -\tmmathbf{r}_c) \cdot
\frac{\partial \hat{H}_c}{\partial \tmmathbf{r}_c} + \frac{\partial
	\hat{H}_c}{\partial \tmmathbf{r}_c} \cdot (\hat{\tmmathbf{r}}
-\tmmathbf{r}_c) \right] \label{Hcorr}.
\end{equation}
 Equation \eqref{Hcorr} follows from the standard Taylor’s expansion, i.e., $\hat{H}[\hat{\bm r},\hat{\bm p};\beta_i(\hat{\bm r})]=\hat{H}[\hat{\bm r},\hat{\bm p};\beta_i(\bm r_c)]+\hat{H}'$, since $\beta(\bm r)$ varies slowly in space.
	
In order to calculate $\ket{\delta u_0}$, the method proposed in Ref.~\cite{Gao2014} is adopted. We construct a wave packet up to first order as
\begin{equation}
|\tilde{W}\rangle = \int d\tmmathbf{k}e^{i\tmmathbf{k} \cdot \tmmathbf{r}}
[C_0 (\tmmathbf{k}) |u_{0\bm k}(\bm r_c)\rangle + \sum_{n \neq 0} C_n (\tmmathbf{k}) |u_{n\bm k}(\bm r_c)
\rangle],\label{eq_wavepacket2}
\end{equation}
where $C_n$ can be determined by requiring the wave packet to satisfy the time-dependent Schr\"{o}dinger
equation $\hat{H} | \tilde{W} \rangle = i \partial_t | \tilde{W} \rangle$ with $\hat{H}=\hat{H}_c+\hat{H}'$. After some lengthy but straightforward calculations~(see Appendix~\ref{Corr} for details), we find
\begin{equation}
C_n  =  \frac{(F_i)_{n 0} [i \partial_{k_i} + (A_{k_i})_{00} - r_{c
		i}]}{\varepsilon_0 - \varepsilon_n} C_0 + \lambda_n C_0,\label{Coeff}
\end{equation}
and
\begin{equation} \label{smallc}
\begin{split}
\lambda_n  &= - \frac{i\partial_{k_i}\vep_0 (F_i)_{n0}}{\varepsilon_0 - \varepsilon_n} + \frac{i \langle u_n |
	\partial_{k_{i}} \hat{F}_i | u_0 \rangle}{2 (\varepsilon_0 - \varepsilon_n)} \\
&\quad +\sum_{m \neq 0}\frac{(F_i)_{nm}(A_{k_i})_{m0}}{\varepsilon_0 - \varepsilon_n},
\end{split}
\end{equation}
where $ \hat{\bm F}=\bm \partial_{\bm r_c}\hat{H}_c$ is the force, $(F_i)_{mn}=\bra{u_{m\mbf{k}}}\hat{F}_i\ket{u_{n\mbf{k}}}$ is its matrix element, $\varepsilon_n$ is the energy of $n$-th band of local Hamiltonian $H_c$, and $(A_{k_i})_{mn}=\bra{u_{m\mbf{k}}}\ket{\partial_{k_{i}}u_{n\mbf{k}}}$ is the Berry connection. In Eq.~(\ref{Coeff}), the first term represents the mixing between adjacent $\mbf{k}$ points within the same band, which is not important in insulators because it only contributes a total derivative of $k_i$ as shown in Eq.~(\ref{RCenter}), whose integration over the entire Brillouin zone vanishes. The second term in Eq.~(\ref{Coeff}) represents mixing between different bands at the same $\bm k$ point~\cite{Gao2015}.  Therefore, in an insulator,
\begin{equation}
\ket{\delta u_0}=\sum_{n\neq 0}\lambda_n\ket{u_n}.\label{CorrWave}
\end{equation}

\subsection{Coarse-grained macroscopic charge density up to second order}

\label{GenB}

In a perfect crystal, the charge density varies drastically on the microscopic scale between neighbouring lattice sites but is uniform on the macroscopic scale much larger than the lattice constant. Here we are concerned with the macroscopic charge density. With the introduction of spatially varying perturbations on the macroscopic scale, we expect the macroscopic charge density to become inhomogeneous.  In this section we will calculate the macroscopic charge density up to second order in spatial gradient in inhomogeneous crystals.

First we need to relate the macroscopic charge density to the microscopic details of the system which are directly calculable from microscopic wave functions.  To this end, we introduce the semiclassical coarse graining procedure based on the wave packet method.  This procedure has been successfully applied to calculate spin density and current density up to first order~\cite{Xiao2006,Culcer2004,Xiao2010}.  Here we show how to calculate the charge density up to second order.

\begin{figure}
	\includegraphics[width=8cm]{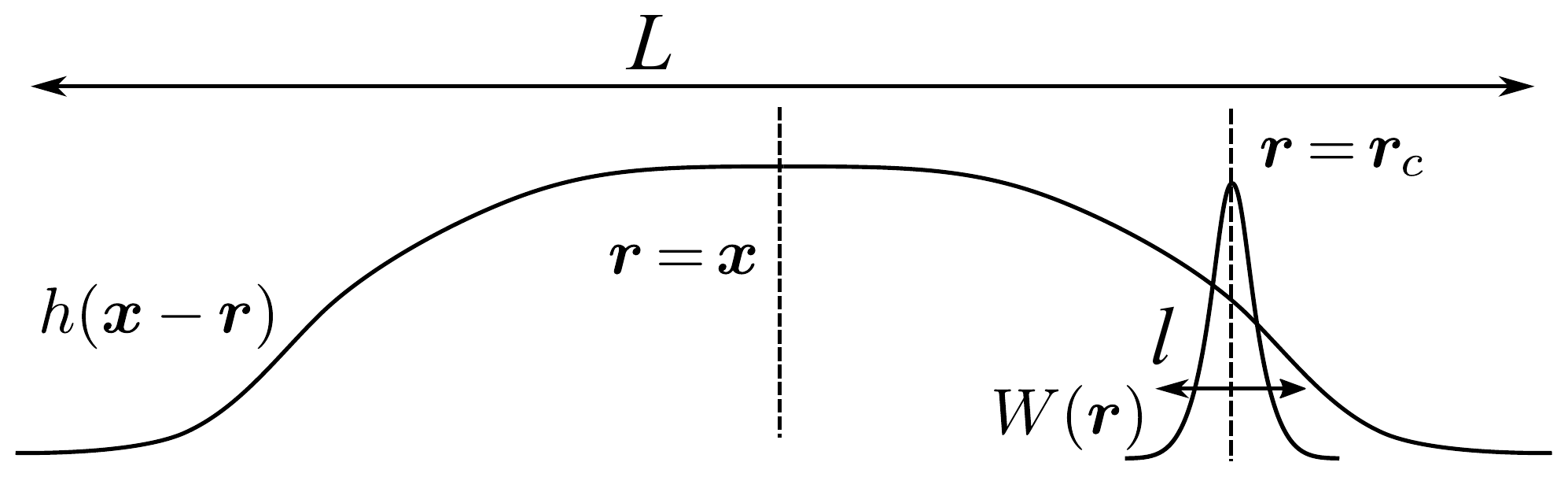}
	\caption{Sampling function $h(\bm x-\mbf{r})$ and wave packet $W(\mbf{r}_c,\mbf{k}_c)$. The width of sampling function $L$ is large compared to wave packet spread $l$ and is small compared to length scale of macroscopic inhomogeneity. Therefore, the sampling function can be approximated by $\delta$ function at the macroscopic level and we can safely perform a Taylor expansion of it within the range of wave packet.}\label{Sampling}
\end{figure}

For simplicity, we consider an insulator at $T=0$ with a single band ($n = 0$) occupied.  We will also set $|e|=1$ throughout this paper.  The charge density can be expressed as follows
\begin{equation}
\rho(\tmmathbf{x}) = -\iint
\frac{d\tmmathbf{r}_c d \tmmathbf{k_c}}{(2 \pi)^d} D (\tmmathbf{r}_c,
\tmmathbf{k}_c) \langle W | \nobracket h (\tmmathbf{x} -
\hat{\bm r}) | W \nobracket \rangle,\label{eq_coarse}
\end{equation}
where $D$ is the modified density of states in Eq.~(\ref{mDos}) and $h (\tmmathbf{x} -
\bm r)$ is a sampling function normalized to unity, i.e., $\int d\mbf{r}\,h(\mbf{x}-\bm r)=1$.  In the above notation, $\bm r$ is the microscopic coordinate, and $\bm x$ is the coarse-grained coordinate.  As shown in Fig.~\ref{Sampling},  $h (\tmmathbf{x} -
\bm r)$ is centered at $\bm r=\mbf{x}$ with a width somewhere between the microscopic scale of the wave packet and the macroscopic scale of the spatial
inhomogeneity. The wave packet $\ket{W(\tmmathbf{r}_c,\tmmathbf{k}_c)}$ hence plays the role of ``molecules'' in the classical coarse graining procedure~\cite{Jackson1999}.

From Eq.~\eqref{eq_coarse}, it is clear that to obtain the charge density, we need two essential elements, i.e., the modified density of states $D$ and the wave packet average of the sampling function. We first calculate $D$.  It is noted from Eq.~(\ref{eq_mDos2}) that $\Omega - J$ is antisymmetric, so $D$ is its Pfaffian. Up to second order we have,
\begin{equation}
\begin{aligned}
D  = & 1 + \tilde{\Omega}_{k_{ci} r_{ci}} -\frac{1}{2}\big(\Omega_{k_{ci}k_{cj}}\Omega_{r_{ci}r_{cj}}\\
& +\Omega_{k_{ci}r_{cj}}\Omega_{k_{cj}r_{ci}}+\Omega_{k_{ci}r_{ci}}\Omega_{r_{cj}k_{cj}}\bigr) \;.\label{mDos2}\\
\end{aligned}
\end{equation}
We emphasize that for the second term, the corrected wave function $\ket{\tilde{u}_0}$ must be used to generate an accurate second-order result, while for the last term, the unperturbed wave function is sufficient because it is already explicitly second order in the spatial gradient.  The second and the last term are the first and second Chern form, respectively.

The third term (the second Chern form) in Eq.~(\ref{mDos2}) is ignored in some previous second-order semi-classical theory~\cite{Gao2014,Gao2015,Gao2018a,Gao2019,Gao2017}  because it vanishes in the special case of uniform electromagnetic fields. To see this, we note that in the case of electric fields, the local Hamiltonian is differed from the unperturbed one by a constant scalar potential and hence the zeroth-order wave function does not depend on the scalar potential and $\bm r_c$, leading to vanishing $\Omega_{r_{ci}r_{cj}}$ and $\Omega_{k_{ci}r_{cj}}$.  In the case of a constant magnetic field $\mbf{B}$, its effect can be taken into account via the Peierls substitution.  Under the symmetric gauge $\mbf{A}=\frac{1}{2}\mbf{B}\times\mbf{r}_c$, the zeroth-order wave function reads $\ket{u_0(\mbf{k}_c+\frac{1}{2}\mbf{B}\times\mbf{r}_c)}$. Therefore, we have
\begin{equation}
\partial_{r_{ci}}=\frac{1}{2}\vep_{ijl}B_l\partial_{k_{cj}}.\label{mag}
\end{equation}
Since the second Chern form is anti-symmetric with respect to all four indices, it has to vanish due to the fact that the Brillouin zone is three-dimensional at most. However, this term can be important in other scenarios. For example, in the case of strain field, it is shown that this term is responsible for the existence of a chiral conducting channel along the line of disclination in metallic systems~\cite{jian2013topological}.

After calculating $D$, next we evaluate the average of the sampling function. We first perform the following expansion
\begin{align}
h (\tmmathbf{x}-\hat{\bm r})&=h [(\tmmathbf{x}-\tmmathbf{r}_c) - \left(\hat{\bm r} -\tmmathbf{r}_c
\right)]\notag\\
&=h(\tmmathbf{x}-\tmmathbf{r}_c)-\frac{\partial h(\bm \xi)}{\partial \xi_i}\bigg|_{\bm \xi=\bm x-\bm r_c}(\hat r_i-r_{ci})\notag\\
&\quad+\frac{1}{2}\frac{\partial^2 h(\bm \xi)}{\partial \xi_i\partial \xi_j}\bigg|_{\bm \xi=\bm x-\bm r_c}(\hat r_i-r_{ci})(\hat r_j-r_{cj})+\cdots.\label{Exp}
\end{align}
This expansion is valid since the sampling function varies slowly within the range of a wave packet. We then approximate the sampling function by the delta function, $h(\mbf{x}-\bm r_c)\approx\delta(\mbf{x}-\bm r_c)$ since its width is much smaller compared to the length scale of the spatial inhomogeneity.

With the help of Eq.~\eqref{Exp}, we can evaluate the average of the sampling function in Eq.~\eqref{eq_coarse} order by order. The zeroth-order term reads
\begin{align}
\bra{W}h(\bm x-\bm r_c)\ket{W}=\delta(\tmmathbf{x}-\tmmathbf{r}_c)\label{eq_zeroth}\,.
 \end{align}
The first-order term vanishes,
\begin{equation} \label{eq_first}
 \begin{split}
&\Bigl\langle W|\frac{\partial h(\bm \xi)}{\partial \xi_i}\bigg|_{\bm \xi=\bm x-\bm r_c}(\hat r_i-r_{ci})|W\Bigr\rangle \\
 =&\frac{\partial h(\bm \xi)}{\partial \xi_i}\bigg|_{\bm \xi=\bm x-\bm r_c}\langle W|\hat r_i-r_{ci}|W\rangle \\
 =&0\,.
\end{split}
\end{equation}
The last equality holds according to the self-consistency condition $\bra{W}\hat{\mbf{r}}\ket{W}=\mbf{r}_c$.  Finally, the second-order term reads (details are left in Appendix~\ref{QuaWave})
\begin{equation}
\begin{split}
&\frac{1}{2}\Bigl\langle W|\frac{\partial^2 h(\bm \xi)}{\partial \xi_i\partial \xi_j}\bigg|_{\bm \xi=\bm x-\bm r_c}(\hat r_i-r_{ci})(\hat r_j-r_{cj})|W\Bigr\rangle \\
=&\frac{1}{2}\frac{\partial^2 \delta(\bm \xi)}{\partial \xi_i\partial \xi_j}\bigg|_{\bm \xi=\bm x-\bm r_c}g_{ij}\label{eq_second}\,.
\end{split}
\end{equation}
Here $g_{ij}$ is the quantum metric tensor of band 0, which can be expressed in terms of the interband Berry connection as follows
\begin{equation}
\begin{aligned}
g_{ij}={\rm Re}\sum_{n\neq 0} (A_{k_{ci}})_{0n} (A_{k_{cj}})_{n0}.
\end{aligned}
\end{equation}
Clearly, $g_{ij}$ has the meaning of the electric quadrupole moment of the wave packet~\cite{Gao2019,Lapa2019}, representing the charge density contribution from its internal structure.  Since Eq.~\eqref{eq_second} is already explicitly second order in spatial derivatives, it is sufficient to use the unperturbed wave function $\ket{u_0}$ and $\ket{u_n}$ in $g_{ij}$.

Plugging Eqs.~\eqref{mDos2}, \eqref{eq_zeroth}--\eqref{eq_second} into Eq.~\eqref{eq_coarse}, we obtain the full expression of the charge density up to second order in spatial gradient,
\begin{equation}\label{eq_rhot}
\rho(\mbf{x})=\rho^{(0)}(\mbf{x})+\rho^{(1)}(\mbf{x})+\rho^{(2)}(\mbf{x}).
\end{equation}
The zeroth-order contribution reads
\begin{align}\label{eq_rho0}
\rho^{(0)}(\mbf{x})=-\frac{1}{V_{\text{cell}}}\,,
\end{align}
where $V_{\text{cell}}$ is the volume of the unit cell, and the minus sign is due to the negative charge carried by electrons. The first-order contribution is
\begin{equation}
\rho^{(1)}(\mbf{x})=-\int_{\text{BZ}}\frac {d \tmmathbf{k}}{(2 \pi)^d} \Omega_{k_i x_i},\label{eq_rho1}
\end{equation}
where the unperturbed wave function $\ket{u_0}$ is used in $\Omega_{k_i x_i}$.  We note that the integration of $\bm r_c$ in Eq.~\eqref{eq_coarse} simply replaces $\bm r_c$ of the integrand with $\bm x$.  Therefore we will use $\bm x$ instead of $\bm r_c$ from now on. We will also drop the subscript $c$ for $\bm k_c$.

Our focus is on the second-order contribution, given by
\begin{equation}
\begin{aligned}
\rho^{(2)}(\mbf{x}) &= \partial_{x_i} \partial_{x_j} q_{ij} - \int_{\text{BZ}}\frac {d \tmmathbf{k}}{(2 \pi)^d} [\delta\Omega_{k_i x_i}\\
& -\frac{1}{2}(\Omega_{k_ik_j}\Omega_{x_ix_j}+\Omega_{k_ix_j}\Omega_{k_jx_i}+\Omega_{k_ix_i}\Omega_{x_jk_j})],\label{eq_rho2}
\end{aligned}
\end{equation}
where
\begin{equation}
q_{ij}=-\int_{\tmop{BZ}} \frac{d
	\tmmathbf{k}}{(2 \pi)^d}\frac{g_{ij}}{2},\label{Quadrupole}
\end{equation}
and $\delta\Omega_{k_i x_i}=i\bra{\partial_{k_i}\delta u_0}\ket{\partial_{x_i}u_0}+i\bra{\partial_{k_i}u_0}\ket{\partial_{x_i}\delta u_0}+c.c.$ is the perturbative correction to the Berry curvature $\Omega_{k_i x_i}$. We note that only the first term (the quadrupole term) in Eq.~\eqref{eq_rho2} is from the spatial average of the sampling function, while the rest comes from the modified density of states in Eq.~(\ref{mDos2}). 

Equation~\eqref{eq_rho2} is the main result of our paper.  The charge density at second order is derived in the most general scenario and hence can be used in diverse cases.  In the following, we will illustrate its meaning and establish its validity.

\subsection{Electric polarization up to first order}

\label{GenC}

The charge density at first and second order in Eqs.~\eqref{eq_rho1} and \eqref{eq_rho2} can be recast in terms of the electric polarization $\mbf{P}$ using Eq.~\eqref{P_density}.  We can divide the electric polarization into different orders in spatial gradient,
\begin{equation}
\bm P=\bm P^{(0)}+\bm P^{(1)}\,,
\end{equation}
corresponding to the first-order and second-order charge density, respectively.

$\bm P^{(0)}$ recovers the familiar result of the electric polarization in a homogeneous system. To see this, we choose the periodic gauge $\ket{\psi_{n\mbf{k}}}=\ket{\psi_{n\mbf{k}+\mbf{G}}}$, where $\mbf{G}$ is the reciprocal lattice vector. Then from Eq.~(\ref{eq_rho1}), we find the following zeroth-order polarization $\mbf{P}^{(0)}$
\begin{equation}
P^{(0)}_i = -\int_{\text{BZ}} \frac{d \tmmathbf{k}}{(2 \pi)^d}
(A_{k_i})_{00}\,.
\end{equation}
This is the exact result originally obtained by King-Smith and Vanderbilt by integrating the adiabatic current~\cite{King-Smith1993}.

We comment that  in the modern theory of the electric polarization from the charge current, it is very important that only the change in the polarization matters, not the polarization itself. To explicitly show this change, artificial time-dependence for the electric polarization is induced, which gives the charge current based on Eq.~\eqref{P_current}. Following the similar logic, here we introduce a spatial dependence of the electric polarization, so that the change of the electric polarization can be reflected. This spatial dependence then translates into the charge density.

Our focus is on the first-order polarization $\mbf{P}^{(1)}$.  It can be divided into three parts: a perturbative part, a topological part, and a quadrupole-like part,
\begin{equation}\label{eq_classp}
P_i^{(1)}=P_i^P+P_i^T+P_i^Q.
\end{equation}
The perturbative part reads
\begin{equation}
P_i^P  =  -\int_{\tmop{BZ}} \frac{d \tmmathbf{k}}{(2 \pi)^d}\label{eq_pertp}
\delta A_{k_i},
\end{equation}
where $\delta A_{k_i}=i\bra{ u_0}\ket{\partial_{k_i}\delta u_0}+c.c.$ is the perturbative correction to the intraband Berry connection $(A_{k_i})_{00}$. This contribution has also been identified in Ref.~\cite{Xiao2009}, but the explicit expression is not given there. From Eqs.~(\ref{smallc}) and (\ref{CorrWave}), we obtain the expression of $\delta A_{k_i}$
\begin{equation}
\begin{aligned}
\delta A_{k_i} & =  \sum_{n \neq 0} \frac{(A_{k_i})_{0 n}}{\varepsilon_0 -
	\varepsilon_n} \bigg[ \sum_{m \neq 0} (F_j)_{nm} (A_{k_j})_{m 0}  + \\
& \quad  +  \frac{i}{2} \langle u_n | \nobracket \partial_{k_j} \hat{F}_j | u_0
\nobracket \rangle - \partial_{k_j} \varepsilon_0  \frac{i (F_j)_{n
		0}}{\varepsilon_0 - \varepsilon_n} \bigg] + c.c..\label{PerturbP}
\end{aligned}
\end{equation}

The topological part $P_i^{T}$ is obtained by evaluating the second Chern form under the periodic gauge, i.e.
\begin{equation}
P_i^T  = -\int_{\text{BZ}} \frac{d \tmmathbf{k_{}}}{(2 \pi)^d} \frac{1}{2}
(A_{x_j} \Omega_{k_i k_j} + A_{k_i} \Omega_{k_j x_j} + A_{k_j}
\Omega_{x_j k_i}).\label{SingleTo}
\end{equation}
We recognize that the integrand in the above equation is of Chern-Simons 3-form. The same expression has also been obtained in Ref.~\cite{Xiao2009}.

The quadrupole-like part $P^Q_i$ comes from the quadrupole moment of the wave packet $g_{ij}$,
\begin{equation}
P^Q_i  =  - \partial_{x_j} q_{ij}.\label{Quadru}
\end{equation}
This is a new term which has not been identified in Ref.~\cite{Xiao2009}.  We will show that this term is significant for the gauge invariance of the first-order electric polarization.

Finally, we mention that when the inhomogeneity is introduced by uniform electromagnetic fields, our result is consistent with previous results~\cite{Gao2014,Essin2010}.  In particular, the quadrupole-like contribution vanishes in these cases.  In the electric field case, the zeroth-order local wave function $\ket{u_0}$ is unchanged, rendering $q_{ij}$ independent of real space coordinate and hence leading to a vanishing $P_i^Q$.  In the case of a constant magnetic field $\mbf{B}$, using Eq.~\eqref{mag} we find that $P_i^Q$ reduces to a total derivative with respect to $\bm k$, whose integration over the entire Brillouin zone has to vanish.

\subsection{Multi-band formulae of the electric polarization}

\label{GenD}

We now generalize our result to the multi-band case.  For the total polarization, this can be done by summing over all occupied bands.  However, in the above we have separated $\bm{P}^{(1)}$ into three contributions.  For this separation to hold physical meanings, each contribution should be invariant under an $U(N)$ gauge transformation in the Hilbert space of occupied bands.  Since the Chern-Simons 3-form and the quantum metric have well known multi-band expressions, we can write down the corresponding polarization in the multi-band case,
\begin{align}
	P^T_i= & -\frac{1}{2}\int_{\text{BZ}}\frac{d\bm{k}}{(2\pi)^d} \tmop{Tr} \left\{ \tmmathbf{A}_{x_j}
	\tmmathbf{\Omega}_{k_i k_j} + \tmmathbf{A}_{k_i}  \tmmathbf{\Omega}_{k_j
		x_j} \right.\nonumber \\
	&  \left.\qquad + \tmmathbf{A}_{k_j}  \tmmathbf{\Omega}_{x_j k_i} + i (\tmmathbf{A}_{x_j}
	\tmmathbf{A}_{k_i} \tmmathbf{A}_{k_j} - \tmmathbf{A}_{x_j}
	\tmmathbf{A}_{k_j} \tmmathbf{A}_{k_i}) \right\}, \label{GaMu} \\
	P_i^Q=&\frac{1}{2}\partial_{x_j}\int_{\text{BZ}}\frac{d\bm{k}}{(2\pi)^d}{\rm Re}\sum_{n\in\text{occ}}^{m\in\text{unocc}} (A_{k_i})_{nm} (A_{k_j})_{mn},\label{eq_multi_quadru}	
\end{align}
where $\tmmathbf{A}_{\xi_i}$ is the matrix form of $(A_{\xi_i})_{nn'}$, $\tmmathbf{\Omega}_{\xi_i\xi_j}=\partial_{\xi_i}\tmmathbf{A}_{\xi_j}-\partial_{\xi_j}\tmmathbf{A}_{\xi_i}-i[\tmmathbf{A}_{\xi_i},\tmmathbf{A}_{\xi_j}]$ is the non-Abelian Berry curvature. 
For the perturbative contribution, the resulting multi-band formula is too complicated (see
Appendix \ref{app_Multi} for details).  We find that it is more convenient to combine $P^Q$ and $P^P$ together into a non-topological contribution $P^N = P^Q + P^P$, which can be written as
\begin{equation}
\begin{aligned}
&\quad P_i^N=\int_{\text{BZ}}\frac{d\bm{k}}{(2\pi)^d}\Re\\
& \sum_{n \in \tmop{occ}}^{m, m' \in \tmop{uno}} \frac{ (V_i)_{n m} (V_j)_{m m'}(F_j)_{m' n} - 
	 (V_i)_{n m}(F_j)_{m m'}(V_j)_{m' n}}{(\varepsilon_n -
	\varepsilon_m)^2 (\varepsilon_n - \varepsilon_{m'})}\\
& + \sum_{n, n' \in \tmop{occ}}^{m \in \tmop{uno}} \frac{(V_i)_{n m}(F_j)_{m n'}(V_j)_{n' n} - 
	(V_i)_{n m} (V_j)_{m n'}(F_j)_{n' n}}{(\varepsilon_n -
	\varepsilon_m)^2 (\varepsilon_{n'} - \varepsilon_m)},\label{NonTo}
\end{aligned}
\end{equation}
where $\hat{V}_i = \partial_{k_i} \hat{H}_c$ is the velocity operator, $(V_i)_{m n} = \langle u_m | \hat{V}_i | u_n \rangle$ is its matrix element. One can readily show that Eq.~\eqref{NonTo} is explicitly gauge invariant. 

We mention that  although the total polarization in the multi-band case is the summation of the single-band polarization over all the occupied bands, each contribution is not. Take the quadrupole-like contribution as an example. By summing over all the occupied bands, it becomes
\begin{equation}
\begin{aligned}
&\frac{1}{2}\partial_{x_j}\int_{\text{BZ}}\frac{d\bm{k}}{(2\pi)^d}{\rm Re}\left[\sum_{n\in\text{occ}}^{m\in\text{unocc}} (A_{k_i})_{nm} (A_{k_j})_{mn}\right.\\
&\qquad\qquad\qquad+\left.\sum_{n,n'\in\text{occ}}^{n\neq n'}(A_{k_i})_{nn'} (A_{k_j})_{n'n}\right].\label{eq_multi_quadru2}
\end{aligned}
\end{equation}
The resulting formula has an additional term which is not gauge invariant compared to Eq.~\eqref{eq_multi_quadru}. Other contributions have similar issues, but the additional terms of all three contributions cancel with each other. Therefore, we can see that the quadrupole-like term from the coarse-graining process plays an important role here, without which one cannot make the total polarization gauge-invariant.

\section{Applications}

\label{Num}

To validate our theory as well as to demonstrate its utility, in this section we apply it to several specific model systems.

\subsection{Strain induced charge density}

We first consider an exactly solvable problem: the charge density in the presence of a constant strain.  The effect of a constant strain is merely a change of the lattice constant from $a$ to $a(1+t)$, and the charge density of the deformed crystal is given by
\begin{equation} \label{eq_exactden}
\begin{split}
\rho_e &= -\frac{1}{a^d(1+t)^d} \\
&= -\frac{1}{V_\text{cell}}\Bigl[1 - dt + \frac{d(d+1)}{2}t^2 + O(t^3)\Bigr] \;,
\end{split}
\end{equation}
where $d$ is the dimension of the system and $V_\text{cell} = a^d$ is the unit cell volume.  We emphasize that only the electron charge density is considered here; the total charge density is always zero due to the charge neutrality condition.

We now derive the charge density using our second-order theory.  A deformed crystal with atomic displacement $\{\bm u_\ell\}$ may be described by the Hamiltonian~\cite{Sundaram1999}
\begin{equation} \label{strainHamiltonian}
\hat{H}=\frac{\mbf{\hat{p}}^2}{2m}+V\big[\mbf{\hat{r}}-\mbf{u}(\hat{\mbf{r}})\big]+s_{ij}(\mbf{\hat{r}})\mathcal{V}_{ij}\big[\mbf{\hat{r}}-\mbf{u}(\mbf{\hat{r}})\big],
\end{equation}
where $V(\bm r)$ is the periodic potential, $\mbf{u}(\mbf{r})$ is the continuous displacement field satisfying $\mbf{u}(\mbf{R}_\ell + \mbf{u}_\ell)=\mbf{u}_\ell$ with $\mbf{R}_\ell$ being the equilibrium position of the $\ell$th atom, and $s_{ij}$ is the unsymmetrized strain tensor $s_{ij}=\partial u_i/\partial r_j$. Detailed derivation of the approximate potential and the definition of $\mathcal{V}_{ij}$ can be found in Appendix~\ref{ApproPoten}.

To apply our theory, the first step is to identify the local Hamiltonian and its first-order correction.  In this case, the local Hamiltonian is obtained by replacing $\bm u(\hat{\bm r})$ with its value at $\bm r_c$ in the full Hamiltonian Eq.~\eqref{strainHamiltonian} and keeping only the zeroth-order term,
\begin{equation}
\hat{H}_c=\frac{\mbf{\hat{p}}^2}{2m}+V[\mbf{\hat{r}}-\mbf{u}(\mbf{r}_c)] \;.
\end{equation}
We see that the effect of a constant displacement field $\bm u(\bm r_c)$ is simply a shift of the position coordinate. Therefore, the periodic part of the local Bloch function is given by $u_{n\mbf{k}}[\mbf{r}-\mbf{u}(\mbf{r}_c)]$.  We caution readers that the continuous displacement field ($\bm{u}$ or $u_i$) should not be confused with the periodic part of the Bloch state, $\ket{u_n}$.

As for the first-order correction, we note that the full Hamiltonian~\eqref{strainHamiltonian} already contains a term that is explicitly first order in the spatial gradient. Therefore the first order correction to the local Hamiltonian contains two terms,
\begin{equation}
\begin{aligned}\label{eq_strainhcorr}
\hat{H}^{(1)} = \hat{H}'+ \delta \hat{H},
\end{aligned}
\end{equation}
where $\hat{H}'$, defined in Eq.~\eqref{Hcorr}, is the gradient expansion of $\hat H_c$, and
\begin{equation}
\delta \hat{H} = s_{ij}(\mbf{r}_c)\mathcal{V}_{ij}\big[\mbf{\hat{r}}-\mbf{u}(\mbf{r}_c)\big].
\end{equation}
 
Next we calculate the Berry connections and Berry curvatures using the unperturbed local Bloch functions.  For simplicity, we assume that only one band is occupied. The Berry connections in the deformed crystal are given by
\begin{align}
A_{r_{ci}}&=i\bra{u_0}\ket{\partial_{r_{ci}}u_0}=f_j(\mbf{k})s_{ji}(\mbf{r}_c), \\
A_{k_{i}}&=i\bra{u_0}\ket{\partial_{k_{i}}u_0} \;,
\end{align}
where $f_i=m \partial_{k_i} \vep_0-k_i$, and we have used the identity $\hat{p}_i=-i\partial_{r_i}=m\partial_{k_i}\hat{H}_c-k_i$.  The corresponding Berry curvatures are 
\begin{equation} \label{eq_berrycur}
\Omega_{r_{ci}r_{cj}}=0, \qquad \Omega_{k_{i}r_{cj}}=s_{lj}\partial_{k_i}f_l.
\end{equation}

With the above preparations, the first-order charge density $\rho^{(1)}$ can be obtained by plugging Eq.~(\ref{eq_berrycur}) into Eq.~(\ref{eq_rho1}),
\begin{equation} \label{first}
\rho^{(1)} =\int_{\text{BZ}}\frac{d\mbf{k}}{(2\pi)^d}s_{ji}\partial_{k_i}f_j
=\frac{s_{ii}}{V_{\text{cell}}}.
\end{equation}

The second-order charge density $\rho^{(2)}$ consists of three parts, $\rho^{(2)}=\rho^Q+\rho^P+\rho^T$.  For the topological part, it is sufficient to use the unperturbed local Bloch functions.  Plugging Eq.~(\ref{eq_berrycur}) into Eq.~(\ref{eq_rho2}), we have 
\begin{equation}
\begin{aligned}
\rho^T&=\frac{1}{2}\int_{\text{BZ}}\frac {d \tmmathbf{k}}{(2 \pi)^d}(s_{lj}\partial_{k_i}f_ls_{ti}\partial_{k_j}f_t-s_{lj}\partial_{k_j}f_ls_{ti}\partial_{k_i}f_t)\\
&=\frac{s_{ij}s_{ji}-s_{ii}s_{jj}}{2V_{\text{cell}}}.
\end{aligned}
\end{equation}
For the quadrupole-like part, it is straightforward to show that $g_{ij}$ is independent of the spatial coordinate, therefore
\begin{equation}
\rho^Q=0.
\end{equation}

The perturbative part has two contributions, arising from corrections to the wave function due to $H'$ and $\delta H$ in Eq.~\eqref{eq_strainhcorr}.  Since $\delta H$ respects the translational symmetry, its correction to the wave function can be readily obtained by perturbation theory.  For $H'$, its contribution to the charge density can be evaluated using Eq.~\eqref{eq_pertp} and \eqref{PerturbP}.  The operator $\hat F$ in Eq.~\eqref{PerturbP} takes the following form in a deformed crystal,
\begin{equation}
\hat{F}_i=-im[\hat{V}_j,\hat{H}_c]s_{ji}\label{eq_strainF}.
\end{equation}
Putting everything together, we arrive at
\begin{equation}
\begin{aligned}
\rho^{P} & =\int_{\text{BZ}}\frac{d \mbf{k}}{(2\pi)^d}  \sum_{n \neq 0} \frac{(A_{k_i})_{0 n}}{\varepsilon_0 -
	\varepsilon_n} \big[(\mathcal{V}_{lj})_{n0}-m\partial_{k_j} \varepsilon_0(V_l)_{n
	0}\\
& +\sum_{n' \neq 0} im(\vep_n-\vep_{n'})(V_l)_{nn'} (A_{k_j})_{n' 0}\big]\frac{\partial s_{lj}}{\partial x_i} + c.c..
\end{aligned}
\end{equation}
where term containing $(\mathcal{V}_{lj})_{n0}=\bra{u_n}\mathcal{V}_{lj}\ket{u_0}$ is due to $\delta \hat{H}$. 

  If a small constant strain is imposed, then $\mbf{u}_\ell = t\mbf{R}_\ell$.  The resulting displacement field and the strain tensor read
\begin{equation}
\mbf{u}(\mbf{x}) = \frac{t}{t+1}\mbf{x}, \quad
s_{ij} = \frac{t}{t+1}\delta_{ij} \;.
\end{equation}
The first-order charge density in Eq.~\eqref{first} becomes
\begin{equation}
\rho^{(1)}=\frac{1}{V_{\text{cell}}}\frac{t}{1+t}d.
\end{equation}
For the second-order charge density, the perturbative contribution $\rho^P$ vanishes since $\partial_{x_i}s_{lj}=0$, and the topological contribution is given by
\begin{equation}
\rho^T=-\frac{1}{V_{\text{cell}}}\frac{d(d-1)}{2}(\frac{t}{t+1})^2 \;.
\end{equation}
We note that $\rho^T=0$ for $d = 1$, because topological part needs at least two dimensions to be nonzero~{\cite{Xiao2009}}.
Adding $\rho^{(0)}$, $\rho^{(1)}$ and $\rho^{(2)}$ together, the charge density up to second order reads
\begin{equation}
\begin{aligned}
\rho&=-\frac{1}{V_{\text{cell}}}\left[1-d\frac{t}{t+1}+\frac{d(d-1)}{2}(\frac{t}{t+1})^2\right]\\
&=-\frac{1}{V_{\text{cell}}}\left[1-dt+\frac{d(d+1)}{2}t^2+O(t^3)\right].\\
\end{aligned}
\end{equation}
This is consistent with the exact result of the charge density in Eq.~(\ref{eq_exactden}), confirming the validity of our theory.

\subsection{Modified SSH model}

In this section, we consider a one-dimensional modified Su-Schrieffer-Heegar (SSH) model.  The focus is on the non-topological contribution of first-order polarization, since the topological contribution vanishes in one-dimensional systems~{\cite{Xiao2009}}. We will also discuss how the coarse graining procedure should be carried out in the numerical simulation.

\begin{figure}
	\centering
	\includegraphics[width=0.9\linewidth]{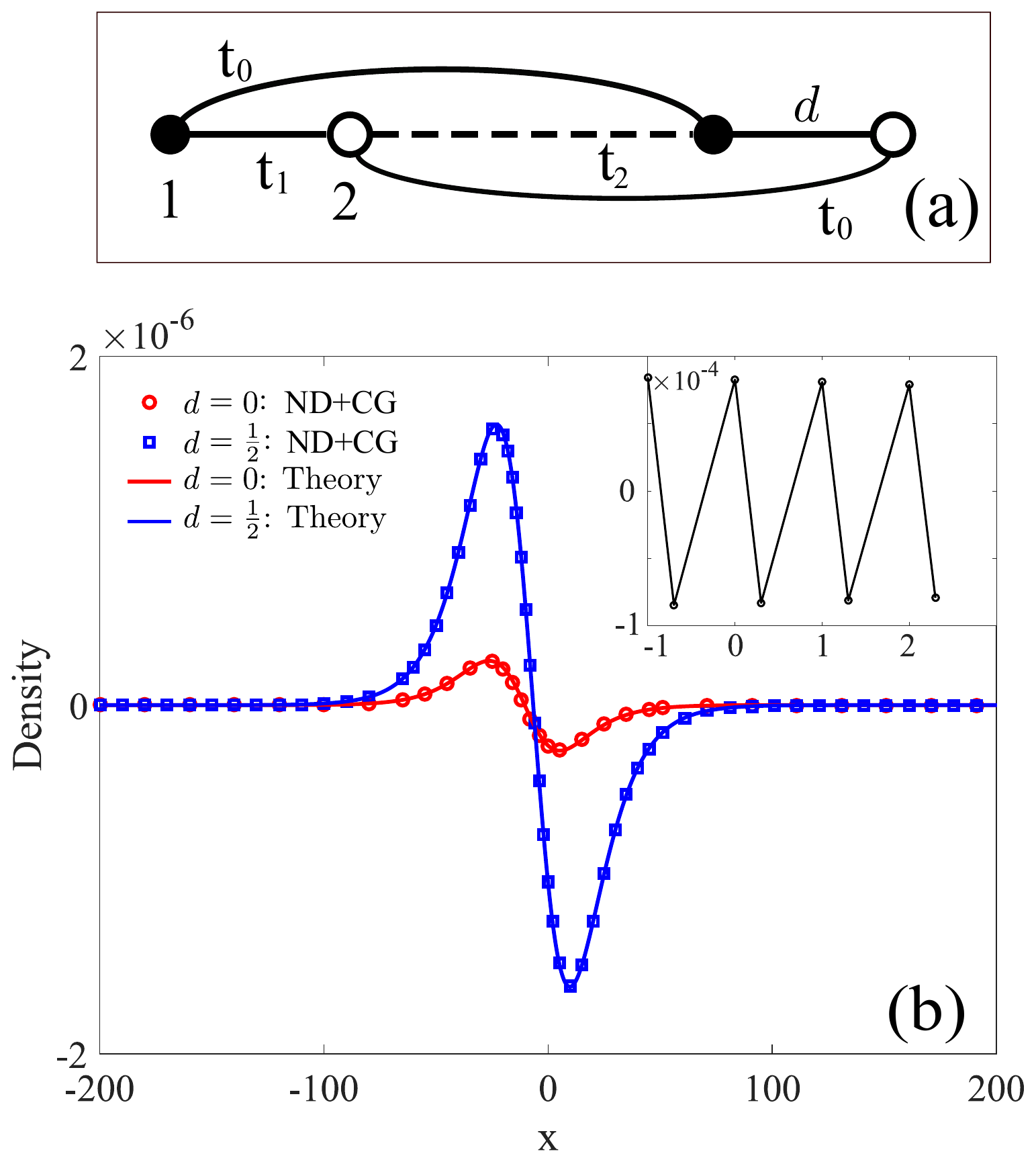}
	\caption{(color online) (a) Model configuration of $H_1$. $t_1$ and $t_2$ are intracell and intercell nearest neighbor hopping, and $t_0$ is second nearest neighbor hopping which breaks particle-hole symmetry. $d$ is the distance between the two sites within the unit cell. (b) Charge density calculated by coarse graining (CG) after numerical diagonalization (ND) and our theory for $d=0, 1/2$. The parameters used in the simulation are: $t_0=0.2,t_1(x)=2+0.3\tanh(x/L), t_2 =1, L=25, \epsilon=2$. The inset of (b) is the charge $q_n^i$ at lattice point calculated by numerical diagonalization of the tight-binding Hamiltonian.}
	\label{model1}
\end{figure}

Our model has two sublattices as depicted in Fig.~\ref{model1}(a) with different hopping strengths $t_1$ and $t_2$. In addition, we add a second nearest neighbor hopping with strength $t_0$ , which makes it different from the original SSH model. The Hamiltonian reads
\begin{equation}
\begin{aligned}
H_1 &= (t_1\hat{a}_{R,1}^\dagger \hat{a}_{R,2}+t_2\hat{a}_{R,1}^\dagger \hat{a}_{R-1,2}+h.c.)\\
&\quad+ t_0(\hat{a}_{R,1}^\dagger \hat{a}_{R+1,1}+\hat{a}_{R,2}^\dagger \hat{a}_{R+1,2}+h.c.), \label{eq_msshr}
\end{aligned}
\end{equation}
where $\hat{a}_{R,i}^\dagger (\hat{a}_{R,i})$ is the electron creation (annihilation) operator on the lattice as shown in Fig.~\ref{model1}(a). The lattice constant is set to be 1.

We introduce the Fourier transformation,
\begin{equation}
\begin{aligned}
\hat{a}_{k,i}&=\frac{1}{\sqrt{N}}\sum_{R}\hat{a}_{R,i}e^{-ik(R+\tau_i)},\\
\hat{a}_{R,i}&=\frac{1}{\sqrt{N}}\sum_{k}\hat{a}_{k,i}e^{ik(R+\tau_i)},
\end{aligned}
\end{equation}
where $\tau_i$ $(i=1,2)$ is the atomic position within the unit cell.  Let $\tau_1=0$ and $\tau_2=d$. Then the Bloch Hamiltonian is
\begin{equation}
\begin{aligned}
H_1 &= 2t_0\cos k\,\sigma_0+[t_1\cos kd+t_2 \cos k(1-d) ]\sigma_x\\
&\quad-[t_1\sin kd-t_2\sin k(1-d)]\sigma_y,\label{eq_mssh}
\end{aligned}
\end{equation}
where $\sigma_x$ and $\sigma_y$ are Pauli matrices in the sublattice space, and $\sigma_0$ is the identity matrix.  It is clear that the second nearest neighbor hopping $t_0$ breaks the particle-hole symmetry.

We now introduce a spatial dependence into $t_1$, with a profile
\begin{equation} \label{t1}
t_1(x)=2+ t\tanh(x/L).
\end{equation}
This inhomogeneity in $t_1$ can induce a polarization. We stress here that the spatial variation of parameters rather than their magnitude must be small for our theory to hold, which means $t$ can be large as long as we keep $t/L$ small.

At zeroth order, the electric polarization depends on the relative strength between $t_1$ and $t_2$~\cite{vanderbilt1993}.  With our choice of $t_1$ in Eq.~\eqref{t1}, we always have $t_1>t_2$ across the entire sample, so $P^{(0)}$ vanishes. Therefore, the leading order contribution to the polarization comes from the first-order contribution.

We now use Eqs.~(\ref{GaMu}) and (\ref{NonTo}) to calculate the first-order polarization and the corresponding charge density.  In one dimension, the topological part of the first-order polarization vanishes~{\cite{Xiao2009}}, so the only nonzero contribution is from the non-topological part $P_i^N$ in Eq.~(\ref{NonTo}).  For $P_i^N$ to be nonzero, the second nearest neighbor hopping is essential because it breaks the particle-hole symmetry.  A detailed discussion can be found in Appendix~\ref{Proof}. The induced polarization reads
\begin{equation}
\begin{aligned}\label{eq_mod1}
P^N&=\int_0^{2\pi}\frac{dk}{2\pi}\left[\frac{t_0t_2^2\sin^2k(t_2+t_1\cos k)}{2(t_1^2+t_2^2+2t_1t_2\cos k)^{5/2}}\right.\\
&\quad\left.-d\frac{t_0t_2\sin^2 k}{2(t_1^2+t_2^2+2t_1t_2\cos k)^{3/2}}\right]\partial_x t_1.\\
\end{aligned}
\end{equation}
The charge density can be obtained by taking the divergence of $P^N$.  We see that the charge density depends on $d$, the distance between the two sites within the unit cell.

To verify our result, we numerically diagonalize the tight-binding Hamiltonian in Eq.~\eqref{eq_msshr} on a finite sample, obtaining the charge at sublattice $i$ of the $n$th unit cell $q_n^i$. Two problems are present here: (\rNum{1}) charge $q_n^i$ oscillates between sublattices as shown in the inset of Fig.~\ref{model1}(b), which is unlikely to produce a smooth charge density; (\rNum{2}) as there is no dependence on intracell site distance $d$ in Eq.~\eqref{eq_msshr}, it is clear that $q_n^i$ is independent of $d$, which seems contradictory to our theory as shown in Eq.~\eqref{eq_mod1}. To reconcile these problems, it is important to keep in mind that our theory gives the macroscopic charge density. Therefore, we have to obtain the numerical macroscopic charge density from the microscopic quantity $q_n^i$. For this purpose, we carry out the coarse graining procedure on the numerical data as follows
\begin{equation}
\begin{aligned}
\rho(x) &= \sum_n [q_n^1 \delta(x-n)+q_n^2 \delta(x-n-d)],\\
\rho_c(x) &= \int dx' h(x-x')\rho(x').
\end{aligned}
\end{equation}
Here $\rho(x)$ is the microscopic charge density, which consists of a series of spikes, and $\rho_c(x)$ is the macroscopic charge density after coarse graining with $h(x)$ being the sampling function as discussed in Sec.~\ref{GenB}.  In our calculation, we have chosen $h(x)=\frac{1}{\sqrt{\pi}\epsilon}\exp(-x^2/\epsilon^2)$.  The coarse-grained charge density $\rho_c(x)$ shows little dependence of $\epsilon$ as long as $\epsilon$ is larger than the lattice constant, but smaller than the length scale of the spatial variation of $t_1(x)$. We can see that in this way the numerical charge density becomes smooth and the $d$-dependence is introduced by $\delta(x-n-d)$. The resulting charge density is plotted in Fig.~\ref{model1}(b) for $d=0$ and $d =\frac{1}{2}$. It is clear that our theory gives excellent agreement in both scenarios.

\subsection{Two dimensional square lattice model}
We now consider a two-dimensional tight-binding model, which has been studied previously in the context of charge fractionalization~\cite{Chamon2008,Seradjeh2008} and higher-order topological insulators~\cite{Benalcazar2017a,Benalcazar2017}. We will focus on the topological contribution of first-order polarization and relate it to the emergence of quantized fractional charge.

\begin{figure}
	\includegraphics[width=8cm]{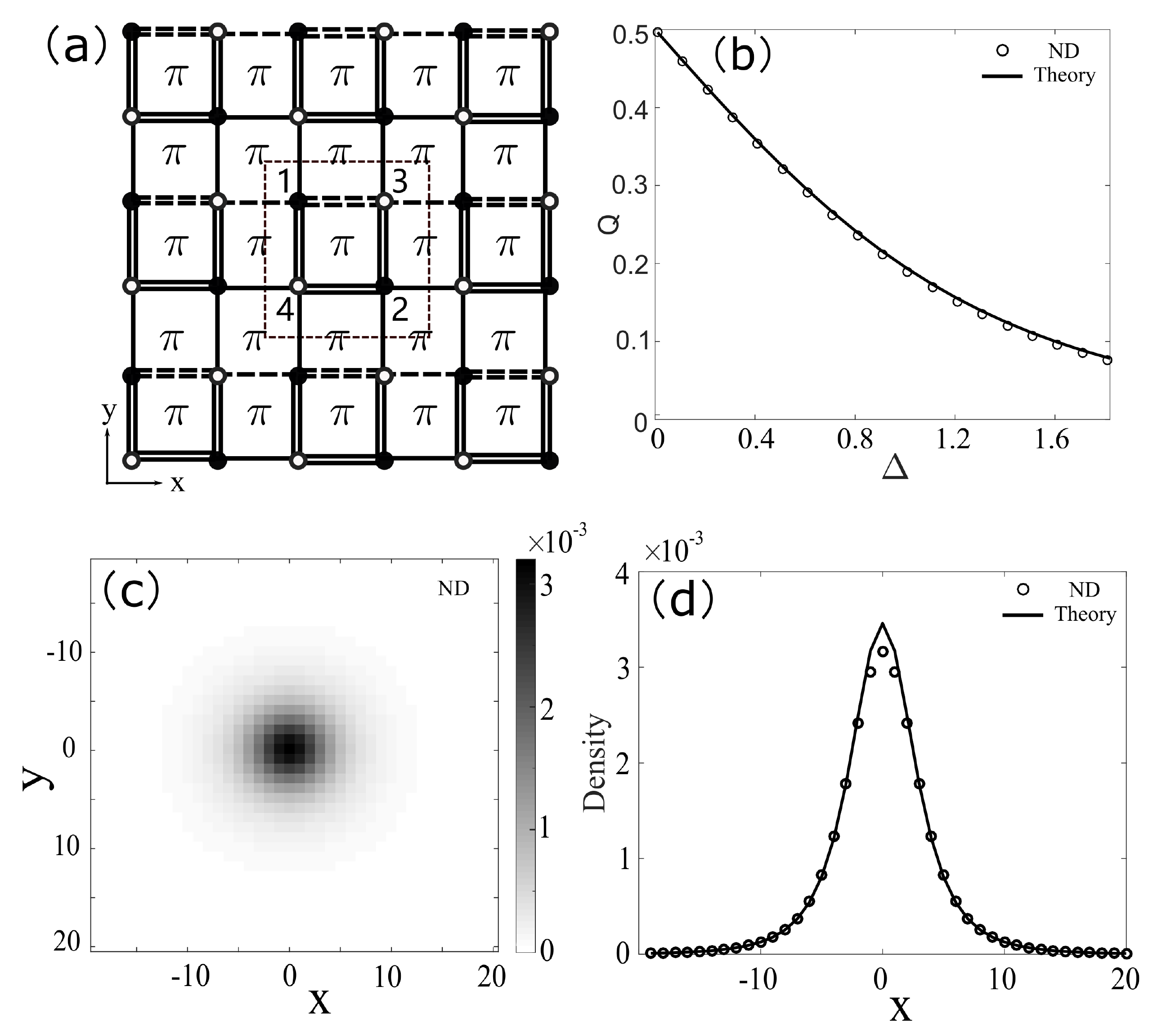}
	\caption{(a) Model configuration of $H_2$. The onsite potential of atoms $1,2$ (atoms $3,4$) is $\Delta$ ($-\Delta$), while dashed lines represent negative signs of hopping resulting from $\pi$ flux threaded through each plaquette. (b) Comparison of total charge calculated by numerical diagonalization (circle) and our theory (solid line). (c) Charge density when $\Delta=0.8$ calculated by numerical diagonalization. (d) Charge density along $y=0$ in (c) calculated by numerical diagonalization (circle) and our theory (solid line). Parameters used in the simulation are $m(r)=0.9 \tanh(r/L), L=10$. The size of the lattice is $40 \times 40$.\label{model2}}
\end{figure}

 As depicted in Fig.~\ref{model2}(a), the model has four atoms in each unit cell forming a square with edge length of $1/2$, while the lattice constants are set to be 1. The onsite potential of atoms $1,2$ (atoms $3,4$) is $\Delta$ ($-\Delta$). The intracell (intercell) hoppings are $1+m_x$ ($1-m_x$) and  $1+m_y$ ($1-m_y$) along the $x$ and $y$ direction, respectively. The dashed line represents a negative sign of the hopping resulting from the $\pi$ flux threading each plaquette. 

The corresponding Bloch Hamiltonian reads
\begin{equation}
\begin{aligned}
H_2=&-2\cos\frac{k_x}{2}\sigma_x\tau_z+2m_x\sin\frac{k_x}{2}\sigma_y\tau_0\\
&+2\cos\frac{k_y}{2}\sigma_x\tau_x+2m_y\sin\frac{k_y}{2}\sigma_x\tau_y\\
&+\Delta\sigma_z\tau_0,
\end{aligned}
\end{equation}
where $\bm{\sigma,\tau}$ are Pauli matrices for the degrees of freedom within a unit cell, and $\tau_0$ is the identity matrix. It has two doubly degenerate bands, with band energies $\pm\vep$,
\begin{equation}
\vep= \sqrt{4\sum_{i=x,y}(\cos^2\frac{k_i}{2}+m_i^2\sin^2\frac{k_i}{2})+\Delta^2}.
\end{equation}
The Hamiltonian is gapped across the whole Brillouin zone unless $m_x=m_y=0$ and $\Delta=0$. We consider the system at half filling, which means the lower doubly degenerate bands are occupied. Suppose there is a vortex in the spatial dependence of $(m_x, m_y)$, i.e.,
\begin{equation}
m_x+i m_y=m(r)e^{i(\theta+\pi/4)},
\end{equation}
where $r,\theta$ are polar coordinates of real space position. We will study the polarization charge carried by the vortex.

At zeroth order, polarization of this model vanishes as long as it is gapped, so the leading order of polarization comes in at the first order. The non-topological first-order  polarization $P_i^N$ vanishes due to the particle-hole symmetry and degeneracy as shown in Appendix~\ref{Proof}. For the topological contribution, it is easier to directly calculate the corresponding charge density in Eq.~\eqref{eq_rho2},
\begin{equation}
\begin{aligned}\label{eq_VortexCharge}
\rho^T&=\int_{\text{BZ}} \frac{d\mbf{k}}{(2\pi)^2}\frac{6\Delta}{\varepsilon^5}\sin^2\frac{k_x}{2}\sin^2\frac{k_y}{2}\frac{m(r)m'(r)}{r},\
\end{aligned}
\end{equation}
which shows that when $|\Delta|$ is small, the charge is concentrated around the the vortex core where $m(r=0)=0,\varepsilon_{min}=|\Delta|$. On the other hand, the parameters $(m_x, m_y)$ vary rapidly near the vortex core, e.g.,
\begin{equation}
\partial_xm_x=\cos\theta\partial_rm_x(r,\theta)-\frac{\sin\theta}{r}\partial_\theta m_x(r,\theta),
\end{equation}
where the second term is divergent at $r=0$. For this reason, our theory can only give the correct charge density away from the vortex core as shown in Fig.~\ref{model2}(d). Fortunately, the total charge can be determined by the polarization at the boundary far from the vortex core, where our theory is valid. The total charge calculated by integration of Eq.~\eqref{eq_VortexCharge} over real space ~(solid line) and diagonalization of tight-binding Hamiltonian~(circle) are plotted in Fig.~\ref{model2}(b), from which we can see that they agree with each other quite well. We note that when $\Delta=0$, the charge carried by the vortex is quantized to $1/2$. Although this quantized fractional charge is already studied in Ref.~\cite{Chamon2008,Seradjeh2008} using a continuum theory, our theory can provide an alternative perspective.

The total charge resulting from the topological part of first-order polarization in two-dimensional systems can be also formulated as
\begin{equation}
\begin{aligned}
Q  &= - \int d{\bm x} \nabla \cdot \bm P^T\\
&=-\int_0^{2\pi} d\theta (rP_r^T)|_{r=+\infty}.
\end{aligned}
\end{equation} 
$P_r^T$ is the radial component of the topological part of first-order polarization, 
\begin{equation}
P_r^T = P_x^T\cos\theta+P_y^T \sin\theta,
\end{equation}
where $P_i^T (i=x,y)$ is the Cartesian component of ${\bm P}^T$ in Eq.~\eqref{SingleTo}.
With the transformation relation between polar and Cartesian coordinate,
\begin{equation}
\begin{aligned}
\partial_{x}= \cos\theta \partial_r-\frac{1}{r}\sin\theta\partial_\theta,\\
\partial_{y}= \sin\theta \partial_r+\frac{1}{r}\cos\theta\partial_\theta,
\end{aligned}
\end{equation}
we can obtain the radial component $P_r^T$,
\begin{equation}
P_r^T =-\frac{1}{2r}\int_{\text{BZ}} \frac{d \tmmathbf{k}}{(2 \pi)^2} (A_{k_x}\Omega_{k_y\theta}+A_{k_y}\Omega_{\theta k_x}+A_{\theta}\Omega_{ k_xk_y}).
\end{equation}
Therefore, the total charge is given by
\begin{equation}
\begin{aligned}
Q = \frac{1}{8\pi^2}\int_0^{2\pi}d\theta\int_{\text{BZ}}d \tmmathbf{k} (A_{k_x}\Omega_{k_y\theta}+A_{k_y}\Omega_{\theta k_x}+A_{\theta}\Omega_{ k_xk_y}),\label{eq_quantizedcharge}
\end{aligned}
\end{equation}
where the integrand is Chern-Simons 3-form in the parameter space $(k_x,k_y,\theta)$. If we treat $\theta$ as the lattice momentum of the third dimension, Eq.~\eqref{eq_quantizedcharge} also gives the quantized magnetoelectric polarizability~\cite{Essin2009} of the effective three-dimensional Hamiltonian $H(k_x,k_y,k_z=\theta)$ without factor $e^2/h$, which is quantized under symmetry reversing the space-time orientation.  Therefore, $Q$ is quantized if the corresponding effective Hamiltonian $H(k_x,k_y,k_z=\theta)$ respects symmetry reversing the space-time orientation. Similar connections are also proposed in Ref.~\cite{Teo2010,Lee2020}.

To understand the quantized fractional charge in our model, we only need to identify the required symmetry. We first define a $C_4$ rotation operator $\hat{r}_4$,
\begin{align}
\hat{r}_4 = \left(\begin{array}{cc}
0 & \tau_x\\
-\tau_z & 0
\end{array}\right),
\end{align}
where $\hat{r}_4$ obeys $\hat{r}_4^4=-1$ (the minus sign is due to the $\pi$ flux per unit cell). Then it can be verified that when $\Delta=0$, 
\begin{equation}
\hat{r}_4H_2(\bm k,\theta)\hat{r}_4^{-1}=H_2(R_4\bm k,-\theta),
\end{equation}
where $R_4$ is the rotation of crystal momentum by $\pi/2$, i.e., $R_4(k_x,k_y)=(k_y,-k_x)$. Alternatively, the corresponding effective three-dimensional Hamiltonian $H_2(k_x,k_y,k_z=\theta)$ has the artificial $C_{4z}M_z$ symmetry \cite{Lee2020}, i.e., $\hat{r}_4H_2(k_x,k_y,k_z)\hat{r}_4^{-1}=H_2(k_y,-k_x,-k_z)$, where $M_z$ is the mirror symmetry with respect to the $xy$-plane. Since $C_{4z}M_z$ symmetry reverses the space-time orientation, the total charge is quantized as shown above.

\section{Summary} \label{Summary}
In this paper, we derive the macroscopic charge density up to second order in spatial gradient in inhomogeneous crystals using semiclassical coarse graining procedure based on the wave packet method. It can be further reformulated by electric polarization, whose first-order contribution consists of a perturbative, a topological and a quadrupole-like part. The topological part can be related to the quantized fractional charge carried by a vortex in two-dimensional systems. Then we generalize our results to gauge-invariant multi-band formulae. Finally, we verify our theory in several model systems. 

\begin{acknowledgments}
 This work is mainly supported by the Department of Energy, Basic Energy Sciences, Grant No.~DE-SC0012509. Y.Z. is also supported by the National Key R\&D Program of China (Grants No.~2017YFA0303302, No.~2018YFA0305602), National Natural Science Foundation of China (Grant No.~11921005). D.X.\ acknowledges the support of a Simons Foundation Fellowship in Theoretical Physics. Y.Z.\ also acknowledges financial support from China Scholarship Council~(No.~201806010045) during his stay at Carnegie Mellon University.
\end{acknowledgments}
\appendix

\section{Derivation of wave function correction up to first order}\label{Corr}

In this section, we will derive the  wave function correction up to first order
in spatial gradient following the method used in Ref.~\cite{Gao2014}. 

For our purpose, we need to construct a wave packet $|\tilde{W} \rangle$ corrected up to first order as shown in Eq.~\eqref{eq_wavepacket2}. As an approximate solution, it should satisfy the time-dependent Schr\"{o}dinger
equation $\hat{H} | \tilde{W} \rangle = i \partial_t | \tilde{W} \rangle$ with $\hat{H}=\hat{H}_c+\hat{H}'$, where $\hat{H}'$ is the gradient expansion of $H_c$ defined in Eq.~\eqref{Hcorr}. We can then use this Schr\"{o}dinger equation to relate expansion coefficient $C_n$ to $C_0$.

We first consider both sides of the Schr\"{o}dinger equation respectively. Since $\ket{u_0} $ and $\ket{u_n}$ all depend on $t$ implicitly through wave packet center $\bm r_c$, the dynamic part of the Schr\"{o}dinger equation reads
\begin{equation}
\begin{aligned}
  & i \partial_t | \tilde{W} \rangle\\
 = & \int d\tmmathbf{k}e^{i\tmmathbf{k} \cdot \tmmathbf{r}} (\tilde{\varepsilon}_0
C_0 |u_0 \rangle + \sum_{n \neq 0} \tilde{\varepsilon}_0 C_n |u_n \rangle)\\
  & + i\int d\tmmathbf{k}e^{i\tmmathbf{k} \cdot \tmmathbf{r}} C_0
\dot{\tmmathbf{r}}_c \cdot \partial_{\tmmathbf{r}_c} |u_0 \rangle\\
  & + i \sum_{n \neq 0} \int d\tmmathbf{k}e^{i\tmmathbf{k} \cdot
	\tmmathbf{r}} C_n \dot{\tmmathbf{r}}_c \cdot \partial_{\tmmathbf{r}_c} |u_n
\rangle,\label{eq_dynamic_S}
\end{aligned}
\end{equation}
where $\tilde{\varepsilon}_0$ is the energy of the wave packet. We have used the identities $i \dot{C_0} = \tilde{\varepsilon}_0$, $i \dot{C}_n = \tilde{\varepsilon}_0$ in the above derivation. The first two terms in Eq.~\eqref{eq_dynamic_S} are the effect of dynamic phase, while the remaining terms result from the change of Bloch states in the parameter space spanned by $\bm r_c$. The energetic part of the Schr\"{o}dinger equation is
\begin{equation}
\begin{aligned}
  & \hat{H} | \tilde{W} \rangle\\
   = & \int d\tmmathbf{k}e^{i\tmmathbf{k} \cdot \tmmathbf{r}} (\varepsilon_0
C_0 |u_0 \rangle + \sum_{n \neq 0} \varepsilon_n C_n |u_n \rangle)\\
  & + \int d\tmmathbf{k}e^{i\tmmathbf{k} \cdot \tmmathbf{r}} (C_0 \hat{H}' |u_0
\rangle + \sum_{n \neq 0} C_n \hat{H}' |u_n \rangle),\label{eq_energy_S}
\end{aligned}
\end{equation}
where $\varepsilon_0$ and $\varepsilon_n$ are the eigenenergies of local Hamiltonian $\hat{H}_c$.

Next we change the integration variables in Eqs.~\eqref{eq_dynamic_S}\eqref{eq_energy_S} from $\bm k$ to $\bm k'$ and take the inner product $\langle u_n |e^{- i\tmmathbf{k} \cdot
\tmmathbf{r}} \nobracket$ to both sides of the Schr\"{o}dinger
equation. In the following derivation, we keep terms up to first order because we only focus on the leading contribution of $C_n$.  The dynamic part is 
\begin{equation}
\begin{aligned}
& \quad  \langle u_n |e^{- i\tmmathbf{k} \cdot \tmmathbf{r}} \nobracket i
\partial_t | W \rangle \\
& =  \tilde{\varepsilon}_0 C_n + C_0 \dot{\tmmathbf{r}}_c \cdot \langle u_n | i
\partial_{\tmmathbf{r}_c} | u_0 \rangle \\
&   \quad + \sum_{m \neq 0} C_m \dot{\tmmathbf{r}}_c \cdot \langle u_n | i
\partial_{\tmmathbf{r}_c} | u_m \rangle \\
& \approx  \varepsilon_0 C_n + C_0 \tmmathbf{v}_0 \cdot \langle u_n | i
\partial_{\tmmathbf{r}_c} | u_0 \rangle, \label{Dy}
\end{aligned}
\end{equation}
where $\tmmathbf{v}_0 = \partial_{\tmmathbf{k}} \varepsilon_0$. In the last step, the term containing $C_m$ is discarded since $C_n$ is of first order in spatial gradient, making this term of second order in total. Wave packet velocity $ \dot{\bm r}_c$ is approximated by band group velocity $\bm v_0$ which is its leading order contribution according to Eq.~\eqref{eq_EOM1}. We also replace the wave packet energy $\tilde{\varepsilon}_0$ with its lowest order contribution $\varepsilon_0$. 

For the energetic part, 
\begin{equation}
\begin{aligned}
&  \quad \langle u_n |e^{- i\tmmathbf{k} \cdot \tmmathbf{r}} \nobracket \hat{H}
| W \rangle \\
& \approx \varepsilon_n  C_n  \label{En}\\
&   + \int d\tmmathbf{k}' C_0 (\tmmathbf{k}') \langle u_{n\bm k} (
\tmmathbf{r}_c) | e^{i (\tmmathbf{k}' -\tmmathbf{k}) \cdot \tmmathbf{r}}
\hat{H}' | u_{0\bm k'} (\tmmathbf{r}_c) \rangle.
\end{aligned}
\end{equation}
The term with both $C_n$ and $\hat{H}'$ is discarded because $\hat{H}'$ is also of first order. We denote the second term of Eq. (\ref{En}) as $\Lambda$. Using the
identity~\cite{Sundaram1999}
\begin{equation}
 \langle \psi_{m\bm k}  | \hat{r}_i | \psi_{n\bm k'} 
\rangle = \left[ (A_{k_i})_{m n} + i \delta_{m n}
\frac{\partial}{\partial k_i} \right] \delta
(\tmmathbf{k}-\tmmathbf{k}'),
\end{equation}
where $(A_{k_i})_{m n} = \langle u_m | i \partial_{k_i}u_n
\rangle$ is the Berry connection, and substituting Eq. (\ref{Hcorr})
into Eq. (\ref{En}), we get
\begin{equation}
\begin{aligned}
\Lambda  = &\frac{1}{2}\sum_m \left[(F_i)_{n m} (A_{k_i})_{m 0}+(A_{k_i})_{n m} (F_i)_{m 0} \right]C_0\\
&+\frac{i}{2} [\partial_{k_i}(F_i)_{n0}]C_0+ (F_i)_{n 0} (i\partial_{k_i} - r_{c i}) C_0,
\end{aligned}
\end{equation}
where operator $\hat{F}_i = \partial_{r_{c i}} \hat{H}_c, (F_i)_{n m} = \langle u_n | \hat{F}_i | u_m{\rangle}$. In the above derivation, we first insert identity $I=\sum_m\int d\bm k\ket{\psi_{m\tmmathbf{k}}}\bra{\psi_{m\tmmathbf{k}}}$ between operator $\hat{\bm F}$ and $\hat{\bm r}$, then use integration by parts. Expanding $\partial_{k_i}F_{n0}$, the above equation becomes 
\begin{equation}
\begin{aligned}
\Lambda  = & \left[\sum_m (F_i)_{n m} (A_{k_i})_{m 0} +\frac{i}{2} \langle u_n | \partial_{k_i} \hat{F}_i | u_0 \rangle\right]C_0\\
  & + (F_i)_{n 0} (i
\partial_{k_i} - r_{c i}) C_0.
\end{aligned}
\end{equation}

With all the preparations, we now compare Eq. (\ref{Dy}) and Eq. (\ref{En}), which gives
\begin{equation}
C_n  =  \frac{(F_i)_{n 0} [i \partial_{k_i} + (A_{k_i})_{00} - r_{c
		i}]}{\varepsilon_0 - \varepsilon_n} C_0 + \lambda_n C_0,
\end{equation}
where
\begin{equation}
\begin{aligned}
\lambda_n  = & - \frac{v_{0 i} \langle u_n | i \partial_{r_{c i}}  u_0
	\rangle}{\varepsilon_0 - \varepsilon_n} + \frac{i \langle u_n |
	\partial_{k_i} F_i | u_0 \rangle}{2 (\varepsilon_0 - \varepsilon_n)}\\
  & + \frac{1}{\varepsilon_0 - \varepsilon_n} \sum_{m \neq 0} (F_i)_{nm}
(A_{k_i})_{m 0}.
\end{aligned}
\end{equation}
Then we can calculate the center of the wave packet
$\tmmathbf{r}_c$ up to first order in spatial gradient,
\begin{equation}
\begin{aligned}
\bm r_c&=   \langle \Psi | \hat{r}_i | \Psi \rangle\\
 &=  \partial_{k_i} \gamma + (A_{k_i})_{00}+\sum_{n \neq 0}[\lambda_n (A_{k_i})_{0 n}+c.c.]\\
  &\quad - \sum_{n \neq 0}  \frac{i}{2} \partial_{k_j}
\left[ \frac{(F_j)_{n 0} (A_{k_i})_{0 n}}{\varepsilon_0 - \varepsilon_n} +c.c.\right] ,\label{RCenter}
\end{aligned}
\end{equation}
where $\gamma$ is the phase of $C_0$. The last term which is total derivative of $k_j$ in the above formula is unimportant in the case of insulator, since its integration over the whole Brillouin zone vanishes. Finally, we come to the conclusion that the first-order correction to the Berry connection of band $0$ is
\begin{equation}
\begin{aligned}
\delta A_{k_i} & =  \sum_{n \neq 0}\lambda_n(A_{k_i})_{0 n} + c.c.,
\end{aligned}
\end{equation}
and the correction to the wave function is 
\begin{equation}
\ket{\delta u_0}=\sum_{n\neq 0}\lambda_n\ket{u_n}.
\end{equation}

\section{Quadrupole moment of wave packet}\label{QuaWave}
In this section we will calculate the quadrupole moment of the wave packet 
\begin{equation}
g_{ij} = \bra{W}(\hat{r}_i-r_{ci})(\hat{r}_j-r_{cj})\ket{W}.
\end{equation}

We first consider the expectation value of operator $\hat{r}_i\hat{r}_j$ on the wave packet $\ket{W}$ 
\begin{equation}
\begin{aligned}
&\quad\bra{W}\hat{r}_i\hat{r}_j\ket{W}\\
&=\iint d\mbf{k}'d\mbf{k}C_0^*(\mbf{k}')C_0(\mbf{k})\bra{u_{0\mbf{k}'}}\partial_{k'_i}e^{-i\mbf{k}'\mbf{r}}\partial_{k_j}e^{i\mbf{k}\mbf{r}}\ket{u_{0\mbf{k}}}.\\
\end{aligned}
\end{equation}
With integration by parts, it becomes
\begin{equation}
\begin{aligned}
&\quad\bra{W}\hat{r}_i\hat{r}_j\ket{W}\\
&=\int d\mbf{k}\partial_{k_i}C_0^*(\mbf{k}')\partial_{k_j}C_0(\mbf{k})\\
&\quad+\int d\mbf{k}\partial_{k_i}C_0^*(\mbf{k})C_0(\mbf{k})\bra{u_{0\mbf{k}}}\ket{\partial_{k_j}u_{0\mbf{k}}}\\
&\quad+\int d\mbf{k}C_0^*(\mbf{k})\partial_{k_j}C_0(\mbf{k})\bra{\partial_{k_i}u_{0\mbf{k}}}\ket{u_{0\mbf{k}}}\\
&\quad+\int d\mbf{k}C_0^*(\mbf{k})C_0(\mbf{k})\bra{\partial_{k_i}u_{0\mbf{k}}}\ket{\partial_{k_j}u_{0\mbf{k}}}.\\
\end{aligned}
\end{equation}
We know that $C_0=|C_0|e^{-i\gamma(\mbf{k})}$ and $|C_0|^2=\delta(\mbf{k}-\mbf{k}_c)$, so it further reduces to 
\begin{equation}
\begin{aligned}
\bra{W}\hat{r}_i\hat{r}_j\ket{W}&=\text{Re}\sum_{n\neq 0}(A_{k_i})_{0n}(A_{k_j})_{n0}+r_{ci}r_{cj},
\end{aligned}
\end{equation}
where wave packet center $r_{ci}=(A_{k_i})_{00}+\partial_{k_i}\gamma$ at the leading order as shown in Eq.~\eqref{RCenter}. Therefore the quadrupole moment of the wave packet is
\begin{equation}
\begin{aligned}
g_{ij}=\text{Re}\sum_{n\neq 0}(A_{k_i})_{0n}(A_{k_j})_{n0}.
\end{aligned}
\end{equation}

\section{Multi-band formulae of electric polarization}\label{app_Multi}

In this section, we will derive the multi-band formulae of the electric polarization by summing up contributions from all the occupied bands. For simplicity, we will omit the integral over lattice momentum $\int_{\text{BZ}}\frac{d\bm{k}}{(2\pi)^d}$ .

We first consider two contributions of the total first-order polarization: quadrupole-like part $P_i^Q$ and perturbative part $P_i^P$.  The quadrupole-like polarization $P_i^Q$ can be reformulated as
\begin{equation}
\begin{aligned}\label{eq_quad}
&\quad P^Q_i\\
&= \frac{1}{4}\sum_{n\neq 0}\left\{\partial_{x_j} \left[ (A_{k_i})_{0 n} (A_{k_j})_{n 0}\right]-\partial_{k_i} \left[ (A_{k_j})_{0 n} (A_{x_j})_{n 0}\right]\right.\\
	&\qquad\qquad\quad\left.+\partial_{k_j} \left[ (A_{x_j})_{0 n} (A_{k_i})_{n 0}\right]\right\}+c.c.\\
	& =-\sum_{n \neq 0}
	\frac{(V_i)_{0 n} (F_j)_{n 0}}{2(\varepsilon_0 - \varepsilon_n)^3} [(V_j)_{0
		0} - (V_j)_{n n}]\\
	&\quad-\sum_{n \neq 0} \frac{(V_i)_{0 n} (V_j)_{n 0}
		[(F_j)_{00} - (F_j)_{n n}]}{2(\varepsilon_0 - \varepsilon_n)^3}\\
	&\quad   + \sum_{n \neq 0}^{m \neq 0, n}  \frac{(V_i)_{0 n} (F_j)_{n m}
		(V_j)_{m 0} + (V_i)_{0 n} (V_j)_{n m} (F_j)_{m 0}}{2(\varepsilon_0 -
		\varepsilon_n)^2 (\varepsilon_0 - \varepsilon_m)} \\
	& \quad + \sum_{n \neq 0} \frac{(V_i)_{0 n} \langle u_n | \partial_{k_j} F_j |
		u_0 \rangle}{2(\varepsilon_0 - \varepsilon_n)^2}+c.c.,
\end{aligned}
\end{equation}
where the integration of the second and third term in the second line over the whole Brillouin zone vanish since they are total derivatives of  $k_i$. In addition, the perturbative part $P_i^P$ [Eq.~\eqref{eq_pertp}] can be transformed into a similar form,
\begin{equation}
\begin{aligned}\label{eq_pertp2}
P_i^P= & \sum_{n \neq 0} \frac{ (V_i)_{0 n} [(F_j)_{n0} (V_j)_{00}-(V_j)_{n 0} (F_j)_{n n}
	]}{(\varepsilon_0 - \varepsilon_n)^3}\\
&- \sum^{m \neq 0, n}_{n \neq
	0} \frac{ (V_i)_{0 n} (F_j)_{n m} (V_j)_{m 0}}{(\varepsilon_0 -
	\varepsilon_n)^2 (\varepsilon_0 - \varepsilon_m)}\\
&  - \sum_{n \neq 0} \frac{(V_i)_{0 n} \langle u_n | \partial_{k_j} F_j |
	u_0 \rangle}{2(\varepsilon_0 - \varepsilon_n)^2}+c.c.. \\
\end{aligned}
\end{equation}

By combining Eqs.~(\ref{eq_quad}) and  (\ref{eq_pertp2}) and generalizing it to multi-band case by summing over all the occupied bands, we have
\begin{equation}
\begin{aligned}\label{eq_quapert}
&\qquad P^Q_i+P_i^P\\
& =\sum_{n\in \text{occ};l\neq n}^{l' \neq n, l}  \frac{(V_i)_{nl} (V_j)_{ll'} (F_j)_{l'n}-(V_i)_{nl} (F_j)_{ll'}
	(V_j)_{l'n}}{2(\varepsilon_n -
	\varepsilon_l)^2 (\varepsilon_n - \varepsilon_{l'})}\\
&\quad+\sum_{n\in \text{occ}}^{l \neq n}
\frac{(V_i)_{nl } (F_j)_{ln}[(V_j)_{nn
	} + (V_j)_{l l}]}{2(\varepsilon_n - \varepsilon_l)^3} \\
&\quad-\sum_{n\in\text{occ}}^{l \neq n} \frac{(V_i)_{nl} (V_j)_{ln}
	[(F_j)_{nn} + (F_j)_{ll}]}{2(\varepsilon_n - \varepsilon_l)^3}+c.c. 
\end{aligned}
\end{equation}
Next, we break the sum over $l,l'$ into contributions from occupied and unoccupied bands. After some manipulations, it can be divided into two parts,
\begin{equation}
\begin{aligned}\label{eq_part1}
& \sum_{n \in \tmop{occ}}^{m, m' \in \tmop{uno}} \frac{ (V_i)_{n m} (V_j)_{m m'}(F_j)_{m' n} - 
	(V_i)_{n m}(F_j)_{m m'}(V_j)_{m' n}}{2(\varepsilon_n -
	\varepsilon_m)^2 (\varepsilon_n - \varepsilon_{m'})}\\
& + \sum_{n, n' \in \tmop{occ}}^{m \in \tmop{uno}} \frac{(V_i)_{n m}(F_j)_{m n'}(V_j)_{n' n} - 
	(V_i)_{n m} (V_j)_{m n'}(F_j)_{n' n}}{2(\varepsilon_n -
	\varepsilon_m)^2 (\varepsilon_{n'} - \varepsilon_m)}\\
&  \qquad\qquad\qquad +c.c.,
\end{aligned}
\end{equation}
and
\begin{equation}
\begin{aligned}\label{eq_part2}
&\quad -\frac{1}{2} \tmop{Tr} \left\{ \tmmathbf{A}_{x_j}
\tmmathbf{\Omega}_{k_i k_j} + \tmmathbf{A}_{k_i}  \tmmathbf{\Omega}_{k_j
	x_j} + \tmmathbf{A}_{k_j}  \tmmathbf{\Omega}_{x_j k_i} \right. \\
&\qquad\qquad\quad\left.+i (\tmmathbf{A}_{x_j}
\tmmathbf{A}_{k_i} \tmmathbf{A}_{k_j} - \tmmathbf{A}_{x_j}
\tmmathbf{A}_{k_j} \tmmathbf{A}_{k_i}) \right\}\\
& \quad+ \frac{1}{2} \sum_{n \in\text{occ}}
[(A_{x_j})_{n n} (\Omega_{k_i k_j})_{n n} + (A_{k_i})_{n n} (\Omega_{k_j
	x_j})_{n n} \\
&\qquad\qquad\qquad+ (A_{k_j})_{n n} (\Omega_{x_j k_i})_{n n}],\\
\end{aligned}
\end{equation}
where $\tmmathbf{A}_{\xi_i}$~($\xi_i\in\{x_i,k_i\}$) is matrix form of $(A_{\xi_i})_{nn'}$, $\tmmathbf{\Omega}_{\xi_i\xi_j}=\partial_{\xi_i}\tmmathbf{A}_{\xi_j}-\partial_{\xi_j}\tmmathbf{A}_{\xi_i}-i[\tmmathbf{A}_{\xi_i},\tmmathbf{A}_{\xi_j}]$ is the non-Abelian Berry curvature matrix. Note that  $(\Omega_{\xi_i\xi_j})_{nn'}=\partial_{\xi_i}(A_{\xi_j})_{nn'}-\partial_{\xi_i}(A_{\xi_j})_{nn'}$ is not the matrix element of $\tmmathbf{\Omega}_{\xi_i\xi_j}$ in the above formula. 

Then we can see that the first part of the Eq.~\eqref{eq_part2} is the standard non-Abelian Chern-Simons 3-form and serves as the natural counterpart of the topological part in multi-band case, while the second part cancels with the multi-band summation of topological part Eq.~(\ref{SingleTo}). It can be shown that the remaining part Eq.~\eqref{eq_part1}, denoted $P_i^N$, is also explicitly gauge invariant.

\section{The approximate potential of strained crystals}\label{ApproPoten}
In this section, we provide a simple derivation of the approximate potential of the strained crystals. A similar but more general derivation can found in Ref.~\cite{Sundaram1999}. 

For simplicity, we assume that the potential of unperturbed crystals is
\begin{equation}
V(\mbf{r})=\sum_{\mbf{R}_\ell}V_0(\mbf{r}-\mbf{R}_{\ell}),
\end{equation}
where $\mbf{R}_{\ell}$ is the lattice vector, $V_0(\mbf{r}-\mbf{R}_{\ell})$ is the local potential around atom at position $\mbf{R}_{\ell}$, which is assumed to decrease sufficiently fast with increasing $|\mbf{r}-\mbf{R}_{\ell}|$. Then the exact potential of strained crystals with atomic displacement $\mbf{u}_{\ell}$ is,
\begin{equation}
\tilde{V}(\mbf{r})=\sum_{\mbf{R}_\ell}V_0(\mbf{r}-\mbf{R}_{\ell}-\mbf{u}_{\ell}).
\end{equation}

To proceed, we can approximate $\mbf{u}_{\ell}$ with continuous displacement field $\mbf{u}(\mbf{r})$,
\begin{equation}
\begin{aligned}
&\quad V_0\big[\mbf{r}-\mbf{R}_{\ell}-\mbf{u}(\mbf{r})+\mbf{u}(\mbf{r})-\mbf{u}_{\ell}\big]\\
&\approx V_0\big[\mbf{r}-\mbf{R}_{\ell}-\mbf{u}(\mbf{r})\big]\\
&\quad+\big[\mbf{u}(\mbf{r})-\mbf{u}_{\ell}\big]\cdot\frac{\partial V_0(\mbf{x})}{\partial\mbf{x}}|_{\mbf{x}=\mbf{r}-\mbf{R}_{\ell}-\mbf{u}(\mbf{r})}.
\end{aligned}
\end{equation}
We require $\mbf{r}=\mbf{R}_{\ell}+\mbf{u}_{\ell}$ to be the zero point of $\mbf{r}-\mbf{R}_{\ell}-\mbf{u}(\mbf{r})=0$ in order to justify the above approximation, so equivalently the atomic displacement $\bm u_\ell$ and continuous displacement field $\bm u(\bm r)$ are related by
\begin{equation}
\mbf{u}(\mbf{R}_{\ell}+\mbf{u}_{\ell})=\mbf{u}_{\ell}.
\end{equation}
Furthermore, $\mbf{u}_{\ell}$ can be approximated by
\begin{equation}
\begin{aligned}
(\mbf{u}_{\ell})_i&=u_i\big(\mbf{r} +\mbf{R}_{\ell}+\mbf{u}_{\ell}-\mbf{r}\big)\\
&\approx u_i(\mbf{r})+\big[(\mbf{R}_\ell)_j+(\mbf{u}_{\ell})_j-r_j\big]s_{ij}(\mbf{r})\\
&\approx u_i(\mbf{r})+\big[(\mbf{R}_\ell)_j+u_j(\mbf{r})-r_j\big]s_{ij}(\mbf{r}),\\
\end{aligned}
\end{equation}
where unsymmetrized strain $s_{ij}=\partial u_i/\partial x_j$. To sum up, the approximate strained potential is
\begin{equation}
\begin{aligned}
\tilde{V}(\mbf{r})\approx V\big[\mbf{r}-\mbf{u}(\mbf{r})\big]+s_{ij}(\mbf{r})\mathcal{V}_{ij}\big[\mbf{r}-\mbf{u}(\mbf{r})\big],
\end{aligned}
\end{equation}
where
\begin{equation}
\mathcal{V}_{ij} = \sum_{\mbf{R}_\ell}\big[r_j-(\mbf{R}_{\ell})_j-u_j(\mbf{r})\big]\frac{\partial V_0(\mbf{x})}{\partial x_i}|_{\mbf{x}=\mbf{r}-\mbf{R}_{\ell}-\mbf{u}(\mbf{r})}.
\end{equation}

\section{A special scenario when non-topological part polarization $P_i^N$ vanishes}\label{Proof}
The special scenario when non-topological part of first-order polarization in Eq.~\eqref{NonTo} vanishes is easily revealed if we formulate it in an alternative form,
\begin{equation}
\begin{aligned}
&\quad P_i^N\\
=& \sum_{n \in \tmop{occ}}^{m, m' \in \tmop{uno}} \frac{i(\varepsilon_m - \varepsilon_{m'})}{2(\varepsilon_n -
	\varepsilon_m)^2 (\varepsilon_n - \varepsilon_{m'})}\left[(F_j)_{m' n}(V_i)_{n m}(A_{k_j})_{m m'}\right. & \\
& \qquad\qquad\left.-(A_{x_j})_{m m'}
(V_j)_{m' n} (V_i)_{n m}\right] & \\
& + \sum_{n, n' \in \tmop{occ}}^{m \in \tmop{uno}}\frac{i (\varepsilon_{n'} - \varepsilon_n)}{2(\varepsilon_n - \varepsilon_m)^2 (\varepsilon_{n'} - \varepsilon_m)}\left[(F_j)_{mn'}(A_{k_j})_{n' n}(V_i)_{n
	m}\right.\\
&\qquad\qquad \left.-(A_{x_j})_{n'
	n}(V_i)_{n m}(V_j)_{m n'}\right]  & \\
& - \sum_{n \in \tmop{occ}} \sum_{m \in \tmop{uno}} \frac{(V_j)_{m n} (V_i)_{n
		m}}{2(\varepsilon_n - \varepsilon_m)^3} \partial_{x_j} (\varepsilon_n +
\varepsilon_m) & \\
& + \sum_{n \in \tmop{occ}} \sum_{m \in \tmop{uno}} \frac{(F_j)_{m n} (V_i)_{n
	m}}{2(\varepsilon_n - \varepsilon_m)^3} \partial_{k_j} (\varepsilon_n +
\varepsilon_m)+c.c.. &
\end{aligned}
\end{equation}
We note that if all the unoccupied (occupied) bands are degenerate, the first (second) term vanishes, and if sum of occupied band energy and unoccupied band energy is constant, the third and fourth term vanish.

To sum up, $P^N$ vanishes identically if the following conditions are satisfied: (\rNum{1}) all the occupied bands are degenerate with energy $E_{\mbf{k}}^v$ at any given momentum $\mbf{k}$; (\rNum{2}) all the unoccupied bands are degenerate with energy $E_{\mbf{k}}^c$ at any given momentum $\mbf{k}$; (\rNum{3}) $E_{\mbf{k}}^v+E_{\mbf{k}}^c=$ constant. A similar discussion is mentioned in Ref.~{\cite{Essin2010}} in the case of magnetic field.


\begin{thebibliography}{57}%
\makeatletter
\providecommand \@ifxundefined [1]{%
 \@ifx{#1\undefined}
}%
\providecommand \@ifnum [1]{%
 \ifnum #1\expandafter \@firstoftwo
 \else \expandafter \@secondoftwo
 \fi
}%
\providecommand \@ifx [1]{%
 \ifx #1\expandafter \@firstoftwo
 \else \expandafter \@secondoftwo
 \fi
}%
\providecommand \natexlab [1]{#1}%
\providecommand \enquote  [1]{``#1''}%
\providecommand \bibnamefont  [1]{#1}%
\providecommand \bibfnamefont [1]{#1}%
\providecommand \citenamefont [1]{#1}%
\providecommand \href@noop [0]{\@secondoftwo}%
\providecommand \href [0]{\begingroup \@sanitize@url \@href}%
\providecommand \@href[1]{\@@startlink{#1}\@@href}%
\providecommand \@@href[1]{\endgroup#1\@@endlink}%
\providecommand \@sanitize@url [0]{\catcode `\\12\catcode `\$12\catcode
  `\&12\catcode `\#12\catcode `\^12\catcode `\_12\catcode `\%12\relax}%
\providecommand \@@startlink[1]{}%
\providecommand \@@endlink[0]{}%
\providecommand \url  [0]{\begingroup\@sanitize@url \@url }%
\providecommand \@url [1]{\endgroup\@href {#1}{\urlprefix }}%
\providecommand \urlprefix  [0]{URL }%
\providecommand \Eprint [0]{\href }%
\providecommand \doibase [0]{http://dx.doi.org/}%
\providecommand \selectlanguage [0]{\@gobble}%
\providecommand \bibinfo  [0]{\@secondoftwo}%
\providecommand \bibfield  [0]{\@secondoftwo}%
\providecommand \translation [1]{[#1]}%
\providecommand \BibitemOpen [0]{}%
\providecommand \bibitemStop [0]{}%
\providecommand \bibitemNoStop [0]{.\EOS\space}%
\providecommand \EOS [0]{\spacefactor3000\relax}%
\providecommand \BibitemShut  [1]{\csname bibitem#1\endcsname}%
\let\auto@bib@innerbib\@empty
\bibitem [{\citenamefont {Jackson}(1999)}]{Jackson1999}%
  \BibitemOpen
  \bibfield  {author} {\bibinfo {author} {\bibfnamefont {J.~D.}\ \bibnamefont
  {Jackson}},\ }\href@noop {} {\emph {\bibinfo {title} {Classical
  Electrodynamics}}},\ \bibinfo {edition} {3rd}\ ed.\ (\bibinfo  {publisher}
  {Wiley},\ \bibinfo {address} {New York},\ \bibinfo {year} {1999})\ pp.\
  \bibinfo {pages} {248--258}\BibitemShut {NoStop}%
\bibitem [{\citenamefont {King-Smith}\ and\ \citenamefont
  {Vanderbilt}(1993)}]{King-Smith1993}%
  \BibitemOpen
  \bibfield  {author} {\bibinfo {author} {\bibfnamefont {R.~D.}\ \bibnamefont
  {King-Smith}}\ and\ \bibinfo {author} {\bibfnamefont {D.}~\bibnamefont
  {Vanderbilt}},\ }\bibfield  {title} {\enquote {\bibinfo {title} {Theory of
  polarization of crystalline solids},}\ }\href {\doibase
  10.1103/PhysRevB.47.1651} {\bibfield  {journal} {\bibinfo  {journal} {Phys.
  Rev. B}\ }\textbf {\bibinfo {volume} {47}},\ \bibinfo {pages} {1651--1654}
  (\bibinfo {year} {1993})}\BibitemShut {NoStop}%
\bibitem [{\citenamefont {Vanderbilt}\ and\ \citenamefont
  {King-Smith}(1993)}]{vanderbilt1993}%
  \BibitemOpen
  \bibfield  {author} {\bibinfo {author} {\bibfnamefont {D.}~\bibnamefont
  {Vanderbilt}}\ and\ \bibinfo {author} {\bibfnamefont {R.~D.}\ \bibnamefont
  {King-Smith}},\ }\bibfield  {title} {\enquote {\bibinfo {title} {Electric
  polarization as a bulk quantity and its relation to surface charge},}\ }\href
  {\doibase 10.1103/PhysRevB.48.4442} {\bibfield  {journal} {\bibinfo
  {journal} {Phys. Rev. B}\ }\textbf {\bibinfo {volume} {48}},\ \bibinfo
  {pages} {4442--4455} (\bibinfo {year} {1993})}\BibitemShut {NoStop}%
\bibitem [{\citenamefont {Resta}(1994)}]{Resta1994}%
  \BibitemOpen
  \bibfield  {author} {\bibinfo {author} {\bibfnamefont {R.}~\bibnamefont
  {Resta}},\ }\bibfield  {title} {\enquote {\bibinfo {title} {Macroscopic
  polarization in crystalline dielectrics: The geometric phase approach},}\
  }\href {\doibase 10.1103/RevModPhys.66.899} {\bibfield  {journal} {\bibinfo
  {journal} {Rev. Mod. Phys.}\ }\textbf {\bibinfo {volume} {66}},\ \bibinfo
  {pages} {899--915} (\bibinfo {year} {1994})}\BibitemShut {NoStop}%
\bibitem [{\citenamefont {Thouless}(1983)}]{Thouless1983}%
  \BibitemOpen
  \bibfield  {author} {\bibinfo {author} {\bibfnamefont {D.~J.}\ \bibnamefont
  {Thouless}},\ }\bibfield  {title} {\enquote {\bibinfo {title} {Quantization
  of particle transport},}\ }\href {\doibase 10.1103/PhysRevB.27.6083}
  {\bibfield  {journal} {\bibinfo  {journal} {Phys. Rev. B}\ }\textbf {\bibinfo
  {volume} {27}},\ \bibinfo {pages} {6083--6087} (\bibinfo {year}
  {1983})}\BibitemShut {NoStop}%
\bibitem [{\citenamefont {Martin}(1972)}]{Martin1972}%
  \BibitemOpen
  \bibfield  {author} {\bibinfo {author} {\bibfnamefont {R.~M.}\ \bibnamefont
  {Martin}},\ }\bibfield  {title} {\enquote {\bibinfo {title}
  {{Piezoelectricity}},}\ }\href {\doibase 10.1103/PhysRevB.5.1607} {\bibfield
  {journal} {\bibinfo  {journal} {Phys. Rev. B}\ }\textbf {\bibinfo {volume}
  {5}},\ \bibinfo {pages} {1607--1613} (\bibinfo {year} {1972})}\BibitemShut
  {NoStop}%
\bibitem [{\citenamefont {Nelson}\ and\ \citenamefont
  {Lax}(1976)}]{Nelson1976}%
  \BibitemOpen
  \bibfield  {author} {\bibinfo {author} {\bibfnamefont {D.~F.}\ \bibnamefont
  {Nelson}}\ and\ \bibinfo {author} {\bibfnamefont {M.}~\bibnamefont {Lax}},\
  }\bibfield  {title} {\enquote {\bibinfo {title} {{Linear elasticity and
  piezoelectricity in pyroelectrics}},}\ }\href {\doibase
  10.1103/PhysRevB.13.1785} {\bibfield  {journal} {\bibinfo  {journal} {Phys.
  Rev. B}\ }\textbf {\bibinfo {volume} {13}},\ \bibinfo {pages} {1785--1796}
  (\bibinfo {year} {1976})}\BibitemShut {NoStop}%
\bibitem [{\citenamefont {{Dal Corso}}\ \emph {et~al.}(1994)\citenamefont {{Dal
  Corso}}, \citenamefont {Posternak}, \citenamefont {Resta},\ and\
  \citenamefont {Baldereschi}}]{DalCorso1994}%
  \BibitemOpen
  \bibfield  {author} {\bibinfo {author} {\bibfnamefont {A.}~\bibnamefont {{Dal
  Corso}}}, \bibinfo {author} {\bibfnamefont {M.}~\bibnamefont {Posternak}},
  \bibinfo {author} {\bibfnamefont {R.}~\bibnamefont {Resta}}, \ and\ \bibinfo
  {author} {\bibfnamefont {A.}~\bibnamefont {Baldereschi}},\ }\bibfield
  {title} {\enquote {\bibinfo {title} {{Ab initio study of piezoelectricity and
  spontaneous polarization in ZnO}},}\ }\href {\doibase
  10.1103/PhysRevB.50.10715} {\bibfield  {journal} {\bibinfo  {journal} {Phys.
  Rev. B}\ }\textbf {\bibinfo {volume} {50}},\ \bibinfo {pages} {10715--10721}
  (\bibinfo {year} {1994})}\BibitemShut {NoStop}%
\bibitem [{\citenamefont {Bernardini}\ \emph {et~al.}(1997)\citenamefont
  {Bernardini}, \citenamefont {Fiorentini},\ and\ \citenamefont
  {Vanderbilt}}]{Bernardini1997}%
  \BibitemOpen
  \bibfield  {author} {\bibinfo {author} {\bibfnamefont {F.}~\bibnamefont
  {Bernardini}}, \bibinfo {author} {\bibfnamefont {V.}~\bibnamefont
  {Fiorentini}}, \ and\ \bibinfo {author} {\bibfnamefont {D.}~\bibnamefont
  {Vanderbilt}},\ }\bibfield  {title} {\enquote {\bibinfo {title} {{Spontaneous
  polarization and piezoelectric constants of III-V nitrides}},}\ }\href
  {\doibase 10.1103/PhysRevB.56.R10024} {\bibfield  {journal} {\bibinfo
  {journal} {Phys. Rev. B}\ }\textbf {\bibinfo {volume} {56}},\ \bibinfo
  {pages} {R10024--R10027} (\bibinfo {year} {1997})}\BibitemShut {NoStop}%
\bibitem [{\citenamefont {S{\'{a}}ghi-Szab{\'{o}}}\ \emph
  {et~al.}(1998)\citenamefont {S{\'{a}}ghi-Szab{\'{o}}}, \citenamefont
  {Cohen},\ and\ \citenamefont {Krakauer}}]{Saghi-Szabo1998}%
  \BibitemOpen
  \bibfield  {author} {\bibinfo {author} {\bibfnamefont {G.}~\bibnamefont
  {S{\'{a}}ghi-Szab{\'{o}}}}, \bibinfo {author} {\bibfnamefont {R.~E.}\
  \bibnamefont {Cohen}}, \ and\ \bibinfo {author} {\bibfnamefont
  {H.}~\bibnamefont {Krakauer}},\ }\bibfield  {title} {\enquote {\bibinfo
  {title} {{First-Principles Study of Piezoelectricity in $\text{PbTiO}_3$}},}\
  }\href {\doibase 10.1103/PhysRevLett.80.4321} {\bibfield  {journal} {\bibinfo
   {journal} {Phys. Rev. Lett.}\ }\textbf {\bibinfo {volume} {80}},\ \bibinfo
  {pages} {4321--4324} (\bibinfo {year} {1998})}\BibitemShut {NoStop}%
\bibitem [{\citenamefont {Vanderbilt}(2000)}]{Vanderbilt2000}%
  \BibitemOpen
  \bibfield  {author} {\bibinfo {author} {\bibfnamefont {D.}~\bibnamefont
  {Vanderbilt}},\ }\bibfield  {title} {\enquote {\bibinfo {title} {{Berry-phase
  theory of proper piezoelectric response}},}\ }\href {\doibase
  10.1016/S0022-3697(99)00273-5} {\bibfield  {journal} {\bibinfo  {journal} {J.
  Phys. Chem. Solids}\ }\textbf {\bibinfo {volume} {61}},\ \bibinfo {pages}
  {147--151} (\bibinfo {year} {2000})}\BibitemShut {NoStop}%
\bibitem [{\citenamefont {Bellaiche}\ and\ \citenamefont
  {Vanderbilt}(2000)}]{Bellaiche2000}%
  \BibitemOpen
  \bibfield  {author} {\bibinfo {author} {\bibfnamefont {L.}~\bibnamefont
  {Bellaiche}}\ and\ \bibinfo {author} {\bibfnamefont {D.}~\bibnamefont
  {Vanderbilt}},\ }\bibfield  {title} {\enquote {\bibinfo {title} {{Virtual
  crystal approximation revisited: Application to dielectric and piezoelectric
  properties of perovskites}},}\ }\href {\doibase 10.1103/PhysRevB.61.7877}
  {\bibfield  {journal} {\bibinfo  {journal} {Phys. Rev. B}\ }\textbf {\bibinfo
  {volume} {61}},\ \bibinfo {pages} {7877--7882} (\bibinfo {year}
  {2000})}\BibitemShut {NoStop}%
\bibitem [{\citenamefont {Liu}\ and\ \citenamefont {Cohen}(2017)}]{Liu2017b}%
  \BibitemOpen
  \bibfield  {author} {\bibinfo {author} {\bibfnamefont {S.}~\bibnamefont
  {Liu}}\ and\ \bibinfo {author} {\bibfnamefont {R.~E.}\ \bibnamefont
  {Cohen}},\ }\bibfield  {title} {\enquote {\bibinfo {title} {{Origin of
  Negative Longitudinal Piezoelectric Effect}},}\ }\href {\doibase
  10.1103/PhysRevLett.119.207601} {\bibfield  {journal} {\bibinfo  {journal}
  {Phys. Rev. Lett.}\ }\textbf {\bibinfo {volume} {119}},\ \bibinfo {pages}
  {207601} (\bibinfo {year} {2017})}\BibitemShut {NoStop}%
\bibitem [{\citenamefont {Hong}\ and\ \citenamefont
  {Vanderbilt}(2013)}]{Hong2013}%
  \BibitemOpen
  \bibfield  {author} {\bibinfo {author} {\bibfnamefont {J.}~\bibnamefont
  {Hong}}\ and\ \bibinfo {author} {\bibfnamefont {D.}~\bibnamefont
  {Vanderbilt}},\ }\bibfield  {title} {\enquote {\bibinfo {title}
  {{First-principles theory and calculation of flexoelectricity}},}\ }\href
  {\doibase 10.1103/PhysRevB.88.174107} {\bibfield  {journal} {\bibinfo
  {journal} {Phys. Rev. B}\ }\textbf {\bibinfo {volume} {88}},\ \bibinfo
  {pages} {174107} (\bibinfo {year} {2013})}\BibitemShut {NoStop}%
\bibitem [{\citenamefont {Resta}(2010)}]{Resta2010}%
  \BibitemOpen
  \bibfield  {author} {\bibinfo {author} {\bibfnamefont {R.}~\bibnamefont
  {Resta}},\ }\bibfield  {title} {\enquote {\bibinfo {title} {{Towards a bulk
  theory of flexoelectricity}},}\ }\href {\doibase
  10.1103/PhysRevLett.105.127601} {\bibfield  {journal} {\bibinfo  {journal}
  {Phys. Rev. Lett.}\ }\textbf {\bibinfo {volume} {105}},\ \bibinfo {pages}
  {127601} (\bibinfo {year} {2010})}\BibitemShut {NoStop}%
\bibitem [{\citenamefont {Hong}\ and\ \citenamefont
  {Vanderbilt}(2011)}]{Hong2011}%
  \BibitemOpen
  \bibfield  {author} {\bibinfo {author} {\bibfnamefont {J.}~\bibnamefont
  {Hong}}\ and\ \bibinfo {author} {\bibfnamefont {D.}~\bibnamefont
  {Vanderbilt}},\ }\bibfield  {title} {\enquote {\bibinfo {title}
  {{First-principles theory of frozen-ion flexoelectricity}},}\ }\href
  {\doibase 10.1103/PhysRevB.84.180101} {\bibfield  {journal} {\bibinfo
  {journal} {Phys. Rev. B}\ }\textbf {\bibinfo {volume} {84}},\ \bibinfo
  {pages} {180101} (\bibinfo {year} {2011})}\BibitemShut {NoStop}%
\bibitem [{\citenamefont {Stengel}(2013)}]{Stengel2013}%
  \BibitemOpen
  \bibfield  {author} {\bibinfo {author} {\bibfnamefont {M.}~\bibnamefont
  {Stengel}},\ }\bibfield  {title} {\enquote {\bibinfo {title}
  {{Flexoelectricity from density-functional perturbation theory}},}\ }\href
  {\doibase 10.1103/PhysRevB.88.174106} {\bibfield  {journal} {\bibinfo
  {journal} {Phys. Rev. B}\ }\textbf {\bibinfo {volume} {88}},\ \bibinfo
  {pages} {174106} (\bibinfo {year} {2013})}\BibitemShut {NoStop}%
\bibitem [{\citenamefont {Schiaffino}\ \emph {et~al.}(2019)\citenamefont
  {Schiaffino}, \citenamefont {Dreyer}, \citenamefont {Vanderbilt},\ and\
  \citenamefont {Stengel}}]{Schiaffino2019}%
  \BibitemOpen
  \bibfield  {author} {\bibinfo {author} {\bibfnamefont {A.}~\bibnamefont
  {Schiaffino}}, \bibinfo {author} {\bibfnamefont {C.~E.}\ \bibnamefont
  {Dreyer}}, \bibinfo {author} {\bibfnamefont {D.}~\bibnamefont {Vanderbilt}},
  \ and\ \bibinfo {author} {\bibfnamefont {M.}~\bibnamefont {Stengel}},\
  }\bibfield  {title} {\enquote {\bibinfo {title} {{Metric wave approach to
  flexoelectricity within density functional perturbation theory}},}\ }\href
  {\doibase 10.1103/PhysRevB.99.085107} {\bibfield  {journal} {\bibinfo
  {journal} {Phys. Rev. B}\ }\textbf {\bibinfo {volume} {99}},\ \bibinfo
  {pages} {085107} (\bibinfo {year} {2019})}\BibitemShut {NoStop}%
\bibitem [{\citenamefont {Essin}\ \emph {et~al.}(2009)\citenamefont {Essin},
  \citenamefont {Moore},\ and\ \citenamefont {Vanderbilt}}]{Essin2009}%
  \BibitemOpen
  \bibfield  {author} {\bibinfo {author} {\bibfnamefont {A.~M.}\ \bibnamefont
  {Essin}}, \bibinfo {author} {\bibfnamefont {J.~E.}\ \bibnamefont {Moore}}, \
  and\ \bibinfo {author} {\bibfnamefont {D.}~\bibnamefont {Vanderbilt}},\
  }\bibfield  {title} {\enquote {\bibinfo {title} {{Magnetoelectric
  Polarizability and Axion Electrodynamics in Crystalline Insulators}},}\
  }\href {\doibase 10.1103/PhysRevLett.102.146805} {\bibfield  {journal}
  {\bibinfo  {journal} {Phys. Rev. Lett.}\ }\textbf {\bibinfo {volume} {102}},\
  \bibinfo {pages} {146805} (\bibinfo {year} {2009})}\BibitemShut {NoStop}%
\bibitem [{\citenamefont {Essin}\ \emph {et~al.}(2010)\citenamefont {Essin},
  \citenamefont {Turner}, \citenamefont {Moore},\ and\ \citenamefont
  {Vanderbilt}}]{Essin2010}%
  \BibitemOpen
  \bibfield  {author} {\bibinfo {author} {\bibfnamefont {A.~M.}\ \bibnamefont
  {Essin}}, \bibinfo {author} {\bibfnamefont {A.~M.}\ \bibnamefont {Turner}},
  \bibinfo {author} {\bibfnamefont {J.~E.}\ \bibnamefont {Moore}}, \ and\
  \bibinfo {author} {\bibfnamefont {D.}~\bibnamefont {Vanderbilt}},\ }\bibfield
   {title} {\enquote {\bibinfo {title} {Orbital magnetoelectric coupling in
  band insulators},}\ }\href {\doibase 10.1103/PhysRevB.81.205104} {\bibfield
  {journal} {\bibinfo  {journal} {Phys. Rev. B}\ }\textbf {\bibinfo {volume}
  {81}},\ \bibinfo {pages} {205104} (\bibinfo {year} {2010})}\BibitemShut
  {NoStop}%
\bibitem [{\citenamefont {Gao}\ \emph {et~al.}(2014)\citenamefont {Gao},
  \citenamefont {Yang},\ and\ \citenamefont {Niu}}]{Gao2014}%
  \BibitemOpen
  \bibfield  {author} {\bibinfo {author} {\bibfnamefont {Y.}~\bibnamefont
  {Gao}}, \bibinfo {author} {\bibfnamefont {S.~A.}\ \bibnamefont {Yang}}, \
  and\ \bibinfo {author} {\bibfnamefont {Q.}~\bibnamefont {Niu}},\ }\bibfield
  {title} {\enquote {\bibinfo {title} {Field induced positional shift of bloch
  electrons and its dynamical implications},}\ }\href {\doibase
  10.1103/PhysRevLett.112.166601} {\bibfield  {journal} {\bibinfo  {journal}
  {Phys. Rev. Lett.}\ }\textbf {\bibinfo {volume} {112}},\ \bibinfo {pages}
  {166601} (\bibinfo {year} {2014})}\BibitemShut {NoStop}%
\bibitem [{\citenamefont {Coh}\ \emph {et~al.}(2011)\citenamefont {Coh},
  \citenamefont {Vanderbilt}, \citenamefont {Malashevich},\ and\ \citenamefont
  {Souza}}]{Coh2011}%
  \BibitemOpen
  \bibfield  {author} {\bibinfo {author} {\bibfnamefont {S.}~\bibnamefont
  {Coh}}, \bibinfo {author} {\bibfnamefont {D.}~\bibnamefont {Vanderbilt}},
  \bibinfo {author} {\bibfnamefont {A.}~\bibnamefont {Malashevich}}, \ and\
  \bibinfo {author} {\bibfnamefont {I.}~\bibnamefont {Souza}},\ }\bibfield
  {title} {\enquote {\bibinfo {title} {{Chern-Simons orbital magnetoelectric
  coupling in generic insulators}},}\ }\href {\doibase
  10.1103/PhysRevB.83.085108} {\bibfield  {journal} {\bibinfo  {journal} {Phys.
  Rev. B}\ }\textbf {\bibinfo {volume} {83}},\ \bibinfo {pages} {085108}
  (\bibinfo {year} {2011})}\BibitemShut {NoStop}%
\bibitem [{\citenamefont {Bousquet}\ \emph {et~al.}(2011)\citenamefont
  {Bousquet}, \citenamefont {Spaldin},\ and\ \citenamefont
  {Delaney}}]{Bousquet2011}%
  \BibitemOpen
  \bibfield  {author} {\bibinfo {author} {\bibfnamefont {E.}~\bibnamefont
  {Bousquet}}, \bibinfo {author} {\bibfnamefont {N.~A.}\ \bibnamefont
  {Spaldin}}, \ and\ \bibinfo {author} {\bibfnamefont {K.~T.}\ \bibnamefont
  {Delaney}},\ }\bibfield  {title} {\enquote {\bibinfo {title} {{Unexpectedly
  Large Electronic Contribution to Linear Magnetoelectricity}},}\ }\href
  {\doibase 10.1103/PhysRevLett.106.107202} {\bibfield  {journal} {\bibinfo
  {journal} {Phys. Rev. Lett.}\ }\textbf {\bibinfo {volume} {106}},\ \bibinfo
  {pages} {107202} (\bibinfo {year} {2011})}\BibitemShut {NoStop}%
\bibitem [{\citenamefont {Malashevich}\ \emph {et~al.}(2012)\citenamefont
  {Malashevich}, \citenamefont {Coh}, \citenamefont {Souza},\ and\
  \citenamefont {Vanderbilt}}]{Malashevich2012}%
  \BibitemOpen
  \bibfield  {author} {\bibinfo {author} {\bibfnamefont {A.}~\bibnamefont
  {Malashevich}}, \bibinfo {author} {\bibfnamefont {S.}~\bibnamefont {Coh}},
  \bibinfo {author} {\bibfnamefont {I.}~\bibnamefont {Souza}}, \ and\ \bibinfo
  {author} {\bibfnamefont {D.}~\bibnamefont {Vanderbilt}},\ }\bibfield  {title}
  {\enquote {\bibinfo {title} {{Full magnetoelectric response of Cr$_2$O$_3$
  from first principles}},}\ }\href {\doibase 10.1103/PhysRevB.86.094430}
  {\bibfield  {journal} {\bibinfo  {journal} {Phys. Rev. B}\ }\textbf {\bibinfo
  {volume} {86}},\ \bibinfo {pages} {094430} (\bibinfo {year}
  {2012})}\BibitemShut {NoStop}%
\bibitem [{\citenamefont {Mostovoy}\ \emph {et~al.}(2010)\citenamefont
  {Mostovoy}, \citenamefont {Scaramucci}, \citenamefont {Spaldin},\ and\
  \citenamefont {Delaney}}]{Mostovoy2010}%
  \BibitemOpen
  \bibfield  {author} {\bibinfo {author} {\bibfnamefont {M.}~\bibnamefont
  {Mostovoy}}, \bibinfo {author} {\bibfnamefont {A.}~\bibnamefont
  {Scaramucci}}, \bibinfo {author} {\bibfnamefont {N.~A.}\ \bibnamefont
  {Spaldin}}, \ and\ \bibinfo {author} {\bibfnamefont {K.~T.}\ \bibnamefont
  {Delaney}},\ }\bibfield  {title} {\enquote {\bibinfo {title}
  {{Temperature-Dependent Magnetoelectric Effect from First Principles}},}\
  }\href {\doibase 10.1103/PhysRevLett.105.087202} {\bibfield  {journal}
  {\bibinfo  {journal} {Phys. Rev. Lett.}\ }\textbf {\bibinfo {volume} {105}},\
  \bibinfo {pages} {087202} (\bibinfo {year} {2010})}\BibitemShut {NoStop}%
\bibitem [{\citenamefont {Malashevich}\ \emph {et~al.}(2010)\citenamefont
  {Malashevich}, \citenamefont {Souza}, \citenamefont {Coh},\ and\
  \citenamefont {Vanderbilt}}]{Malashevich2010}%
  \BibitemOpen
  \bibfield  {author} {\bibinfo {author} {\bibfnamefont {A.}~\bibnamefont
  {Malashevich}}, \bibinfo {author} {\bibfnamefont {I.}~\bibnamefont {Souza}},
  \bibinfo {author} {\bibfnamefont {S.}~\bibnamefont {Coh}}, \ and\ \bibinfo
  {author} {\bibfnamefont {D.}~\bibnamefont {Vanderbilt}},\ }\bibfield  {title}
  {\enquote {\bibinfo {title} {{Theory of orbital magnetoelectric response}},}\
  }\href {\doibase 10.1088/1367-2630/12/5/053032} {\bibfield  {journal}
  {\bibinfo  {journal} {New J. Phys.}\ }\textbf {\bibinfo {volume} {12}},\
  \bibinfo {pages} {053032} (\bibinfo {year} {2010})}\BibitemShut {NoStop}%
\bibitem [{\citenamefont {Lawes}\ \emph {et~al.}(2005)\citenamefont {Lawes},
  \citenamefont {Harris}, \citenamefont {Kimura}, \citenamefont {Rogado},
  \citenamefont {Cava}, \citenamefont {Aharony}, \citenamefont {Entin-Wohlman},
  \citenamefont {Yildirim}, \citenamefont {Kenzelmann}, \citenamefont
  {Broholm},\ and\ \citenamefont {Ramirez}}]{Lawes2005}%
  \BibitemOpen
  \bibfield  {author} {\bibinfo {author} {\bibfnamefont {G.}~\bibnamefont
  {Lawes}}, \bibinfo {author} {\bibfnamefont {A.~B.}\ \bibnamefont {Harris}},
  \bibinfo {author} {\bibfnamefont {T.}~\bibnamefont {Kimura}}, \bibinfo
  {author} {\bibfnamefont {N.}~\bibnamefont {Rogado}}, \bibinfo {author}
  {\bibfnamefont {R.~J.}\ \bibnamefont {Cava}}, \bibinfo {author}
  {\bibfnamefont {A.}~\bibnamefont {Aharony}}, \bibinfo {author} {\bibfnamefont
  {O.}~\bibnamefont {Entin-Wohlman}}, \bibinfo {author} {\bibfnamefont
  {T.}~\bibnamefont {Yildirim}}, \bibinfo {author} {\bibfnamefont
  {M.}~\bibnamefont {Kenzelmann}}, \bibinfo {author} {\bibfnamefont
  {C.}~\bibnamefont {Broholm}}, \ and\ \bibinfo {author} {\bibfnamefont
  {A.~P.}\ \bibnamefont {Ramirez}},\ }\bibfield  {title} {\enquote {\bibinfo
  {title} {Magnetically driven ferroelectric order in
  $\text{Ni}_3\text{V}_2\text{O}_8$},}\ }\href {\doibase
  10.1103/PhysRevLett.95.087205} {\bibfield  {journal} {\bibinfo  {journal}
  {Phys. Rev. Lett.}\ }\textbf {\bibinfo {volume} {95}},\ \bibinfo {pages}
  {087205} (\bibinfo {year} {2005})}\BibitemShut {NoStop}%
\bibitem [{\citenamefont {Kenzelmann}\ \emph {et~al.}(2005)\citenamefont
  {Kenzelmann}, \citenamefont {Harris}, \citenamefont {Jonas}, \citenamefont
  {Broholm}, \citenamefont {Schefer}, \citenamefont {Kim}, \citenamefont
  {Zhang}, \citenamefont {Cheong}, \citenamefont {Vajk},\ and\ \citenamefont
  {Lynn}}]{Kenzelmann2005}%
  \BibitemOpen
  \bibfield  {author} {\bibinfo {author} {\bibfnamefont {M.}~\bibnamefont
  {Kenzelmann}}, \bibinfo {author} {\bibfnamefont {A.~B.}\ \bibnamefont
  {Harris}}, \bibinfo {author} {\bibfnamefont {S.}~\bibnamefont {Jonas}},
  \bibinfo {author} {\bibfnamefont {C.}~\bibnamefont {Broholm}}, \bibinfo
  {author} {\bibfnamefont {J.}~\bibnamefont {Schefer}}, \bibinfo {author}
  {\bibfnamefont {S.~B.}\ \bibnamefont {Kim}}, \bibinfo {author} {\bibfnamefont
  {C.~L.}\ \bibnamefont {Zhang}}, \bibinfo {author} {\bibfnamefont {S.-W.}\
  \bibnamefont {Cheong}}, \bibinfo {author} {\bibfnamefont {O.~P.}\
  \bibnamefont {Vajk}}, \ and\ \bibinfo {author} {\bibfnamefont {J.~W.}\
  \bibnamefont {Lynn}},\ }\bibfield  {title} {\enquote {\bibinfo {title}
  {Magnetic inversion symmetry breaking and ferroelectricity in
  $\text{TbMnO}_3$},}\ }\href {\doibase 10.1103/PhysRevLett.95.087206}
  {\bibfield  {journal} {\bibinfo  {journal} {Phys. Rev. Lett.}\ }\textbf
  {\bibinfo {volume} {95}},\ \bibinfo {pages} {087206} (\bibinfo {year}
  {2005})}\BibitemShut {NoStop}%
\bibitem [{\citenamefont {Neaton}\ \emph {et~al.}(2005)\citenamefont {Neaton},
  \citenamefont {Ederer}, \citenamefont {Waghmare}, \citenamefont {Spaldin},\
  and\ \citenamefont {Rabe}}]{Neaton2005}%
  \BibitemOpen
  \bibfield  {author} {\bibinfo {author} {\bibfnamefont {J.~B.}\ \bibnamefont
  {Neaton}}, \bibinfo {author} {\bibfnamefont {C.}~\bibnamefont {Ederer}},
  \bibinfo {author} {\bibfnamefont {U.~V.}\ \bibnamefont {Waghmare}}, \bibinfo
  {author} {\bibfnamefont {N.~A.}\ \bibnamefont {Spaldin}}, \ and\ \bibinfo
  {author} {\bibfnamefont {K.~M.}\ \bibnamefont {Rabe}},\ }\bibfield  {title}
  {\enquote {\bibinfo {title} {{First-principles study of spontaneous
  polarization in multiferroic BiFeO$_3$}},}\ }\href {\doibase
  10.1103/PhysRevB.71.014113} {\bibfield  {journal} {\bibinfo  {journal} {Phys.
  Rev. B}\ }\textbf {\bibinfo {volume} {71}},\ \bibinfo {pages} {014113}
  (\bibinfo {year} {2005})}\BibitemShut {NoStop}%
\bibitem [{\citenamefont {Katsura}\ \emph {et~al.}(2005)\citenamefont
  {Katsura}, \citenamefont {Nagaosa},\ and\ \citenamefont
  {Balatsky}}]{Katsura2005}%
  \BibitemOpen
  \bibfield  {author} {\bibinfo {author} {\bibfnamefont {H.}~\bibnamefont
  {Katsura}}, \bibinfo {author} {\bibfnamefont {N.}~\bibnamefont {Nagaosa}}, \
  and\ \bibinfo {author} {\bibfnamefont {A.~V.}\ \bibnamefont {Balatsky}},\
  }\bibfield  {title} {\enquote {\bibinfo {title} {{Spin Current and
  Magnetoelectric Effect in Noncollinear Magnets}},}\ }\href {\doibase
  10.1103/PhysRevLett.95.057205} {\bibfield  {journal} {\bibinfo  {journal}
  {Phys. Rev. Lett.}\ }\textbf {\bibinfo {volume} {95}},\ \bibinfo {pages}
  {057205} (\bibinfo {year} {2005})}\BibitemShut {NoStop}%
\bibitem [{\citenamefont {Jia}\ \emph {et~al.}(2006)\citenamefont {Jia},
  \citenamefont {Onoda}, \citenamefont {Nagaosa},\ and\ \citenamefont
  {Han}}]{Jia2006}%
  \BibitemOpen
  \bibfield  {author} {\bibinfo {author} {\bibfnamefont {C.}~\bibnamefont
  {Jia}}, \bibinfo {author} {\bibfnamefont {S.}~\bibnamefont {Onoda}}, \bibinfo
  {author} {\bibfnamefont {N.}~\bibnamefont {Nagaosa}}, \ and\ \bibinfo
  {author} {\bibfnamefont {J.~H.}\ \bibnamefont {Han}},\ }\bibfield  {title}
  {\enquote {\bibinfo {title} {{Bond electronic polarization induced by
  spin}},}\ }\href {\doibase 10.1103/PhysRevB.74.224444} {\bibfield  {journal}
  {\bibinfo  {journal} {Phys. Rev. B}\ }\textbf {\bibinfo {volume} {74}},\
  \bibinfo {pages} {224444} (\bibinfo {year} {2006})}\BibitemShut {NoStop}%
\bibitem [{\citenamefont {Mostovoy}(2006)}]{Mostovoy2006}%
  \BibitemOpen
  \bibfield  {author} {\bibinfo {author} {\bibfnamefont {M.}~\bibnamefont
  {Mostovoy}},\ }\bibfield  {title} {\enquote {\bibinfo {title}
  {{Ferroelectricity in Spiral Magnets}},}\ }\href {\doibase
  10.1103/PhysRevLett.96.067601} {\bibfield  {journal} {\bibinfo  {journal}
  {Phys. Rev. Lett.}\ }\textbf {\bibinfo {volume} {96}},\ \bibinfo {pages}
  {067601} (\bibinfo {year} {2006})}\BibitemShut {NoStop}%
\bibitem [{\citenamefont {Jia}\ \emph {et~al.}(2007)\citenamefont {Jia},
  \citenamefont {Onoda}, \citenamefont {Nagaosa},\ and\ \citenamefont
  {Han}}]{Jia2007}%
  \BibitemOpen
  \bibfield  {author} {\bibinfo {author} {\bibfnamefont {C.}~\bibnamefont
  {Jia}}, \bibinfo {author} {\bibfnamefont {S.}~\bibnamefont {Onoda}}, \bibinfo
  {author} {\bibfnamefont {N.}~\bibnamefont {Nagaosa}}, \ and\ \bibinfo
  {author} {\bibfnamefont {J.~H.}\ \bibnamefont {Han}},\ }\bibfield  {title}
  {\enquote {\bibinfo {title} {{Microscopic theory of spin-polarization
  coupling in multiferroic transition metal oxides}},}\ }\href {\doibase
  10.1103/PhysRevB.76.144424} {\bibfield  {journal} {\bibinfo  {journal} {Phys.
  Rev. B}\ }\textbf {\bibinfo {volume} {76}},\ \bibinfo {pages} {144424}
  (\bibinfo {year} {2007})}\BibitemShut {NoStop}%
\bibitem [{\citenamefont {Harris}(2007)}]{Harris2007}%
  \BibitemOpen
  \bibfield  {author} {\bibinfo {author} {\bibfnamefont {A.~B.}\ \bibnamefont
  {Harris}},\ }\bibfield  {title} {\enquote {\bibinfo {title} {{Landau analysis
  of the symmetry of the magnetic structure and magnetoelectric interaction in
  multiferroics}},}\ }\href {\doibase 10.1103/PhysRevB.76.054447} {\bibfield
  {journal} {\bibinfo  {journal} {Phys. Rev. B}\ }\textbf {\bibinfo {volume}
  {76}},\ \bibinfo {pages} {054447} (\bibinfo {year} {2007})}\BibitemShut
  {NoStop}%
\bibitem [{\citenamefont {Kenzelmann}\ \emph {et~al.}(2007)\citenamefont
  {Kenzelmann}, \citenamefont {Lawes}, \citenamefont {Harris}, \citenamefont
  {Gasparovic}, \citenamefont {Broholm}, \citenamefont {Ramirez}, \citenamefont
  {Jorge}, \citenamefont {Jaime}, \citenamefont {Park}, \citenamefont {Huang},
  \citenamefont {Shapiro},\ and\ \citenamefont {Demianets}}]{Kenzelmann2007}%
  \BibitemOpen
  \bibfield  {author} {\bibinfo {author} {\bibfnamefont {M.}~\bibnamefont
  {Kenzelmann}}, \bibinfo {author} {\bibfnamefont {G.}~\bibnamefont {Lawes}},
  \bibinfo {author} {\bibfnamefont {A.~B.}\ \bibnamefont {Harris}}, \bibinfo
  {author} {\bibfnamefont {G.}~\bibnamefont {Gasparovic}}, \bibinfo {author}
  {\bibfnamefont {C.}~\bibnamefont {Broholm}}, \bibinfo {author} {\bibfnamefont
  {A.~P.}\ \bibnamefont {Ramirez}}, \bibinfo {author} {\bibfnamefont {G.~A.}\
  \bibnamefont {Jorge}}, \bibinfo {author} {\bibfnamefont {M.}~\bibnamefont
  {Jaime}}, \bibinfo {author} {\bibfnamefont {S.}~\bibnamefont {Park}},
  \bibinfo {author} {\bibfnamefont {Q.}~\bibnamefont {Huang}}, \bibinfo
  {author} {\bibfnamefont {A.~Y.}\ \bibnamefont {Shapiro}}, \ and\ \bibinfo
  {author} {\bibfnamefont {L.~A.}\ \bibnamefont {Demianets}},\ }\bibfield
  {title} {\enquote {\bibinfo {title} {{Direct Transition from a Disordered to
  a Multiferroic Phase on a Triangular Lattice}},}\ }\href {\doibase
  10.1103/PhysRevLett.98.267205} {\bibfield  {journal} {\bibinfo  {journal}
  {Phys. Rev. Lett.}\ }\textbf {\bibinfo {volume} {98}},\ \bibinfo {pages}
  {267205} (\bibinfo {year} {2007})}\BibitemShut {NoStop}%
\bibitem [{\citenamefont {Malashevich}\ and\ \citenamefont
  {Vanderbilt}(2008)}]{Malashevich2008}%
  \BibitemOpen
  \bibfield  {author} {\bibinfo {author} {\bibfnamefont {A.}~\bibnamefont
  {Malashevich}}\ and\ \bibinfo {author} {\bibfnamefont {D.}~\bibnamefont
  {Vanderbilt}},\ }\bibfield  {title} {\enquote {\bibinfo {title} {First
  principles study of improper ferroelectricity in $\text{TbMnO}_3$},}\ }\href
  {\doibase 10.1103/PhysRevLett.101.037210} {\bibfield  {journal} {\bibinfo
  {journal} {Phys. Rev. Lett.}\ }\textbf {\bibinfo {volume} {101}},\ \bibinfo
  {pages} {037210} (\bibinfo {year} {2008})}\BibitemShut {NoStop}%
\bibitem [{\citenamefont {Malashevich}\ and\ \citenamefont
  {Vanderbilt}(2009)}]{Malashevich2009}%
  \BibitemOpen
  \bibfield  {author} {\bibinfo {author} {\bibfnamefont {A.}~\bibnamefont
  {Malashevich}}\ and\ \bibinfo {author} {\bibfnamefont {D.}~\bibnamefont
  {Vanderbilt}},\ }\bibfield  {title} {\enquote {\bibinfo {title} {{Dependence
  of electronic polarization on octahedral rotations in TbMnO3 from first
  principles}},}\ }\href {\doibase 10.1103/PhysRevB.80.224407} {\bibfield
  {journal} {\bibinfo  {journal} {Phys. Rev. B}\ }\textbf {\bibinfo {volume}
  {80}},\ \bibinfo {pages} {224407} (\bibinfo {year} {2009})}\BibitemShut
  {NoStop}%
\bibitem [{\citenamefont {Xiang}\ \emph {et~al.}(2011)\citenamefont {Xiang},
  \citenamefont {Kan}, \citenamefont {Zhang}, \citenamefont {Whangbo},\ and\
  \citenamefont {Gong}}]{Xiang2011}%
  \BibitemOpen
  \bibfield  {author} {\bibinfo {author} {\bibfnamefont {H.~J.}\ \bibnamefont
  {Xiang}}, \bibinfo {author} {\bibfnamefont {E.~J.}\ \bibnamefont {Kan}},
  \bibinfo {author} {\bibfnamefont {Y.}~\bibnamefont {Zhang}}, \bibinfo
  {author} {\bibfnamefont {M.-H.}\ \bibnamefont {Whangbo}}, \ and\ \bibinfo
  {author} {\bibfnamefont {X.~G.}\ \bibnamefont {Gong}},\ }\bibfield  {title}
  {\enquote {\bibinfo {title} {{General Theory for the Ferroelectric
  Polarization Induced by Spin-Spiral Order}},}\ }\href {\doibase
  10.1103/PhysRevLett.107.157202} {\bibfield  {journal} {\bibinfo  {journal}
  {Phys. Rev. Lett.}\ }\textbf {\bibinfo {volume} {107}},\ \bibinfo {pages}
  {157202} (\bibinfo {year} {2011})}\BibitemShut {NoStop}%
\bibitem [{\citenamefont {Xiang}\ \emph {et~al.}(2013)\citenamefont {Xiang},
  \citenamefont {Wang}, \citenamefont {Whangbo},\ and\ \citenamefont
  {Gong}}]{Xiang2013}%
  \BibitemOpen
  \bibfield  {author} {\bibinfo {author} {\bibfnamefont {H.~J.}\ \bibnamefont
  {Xiang}}, \bibinfo {author} {\bibfnamefont {P.~S.}\ \bibnamefont {Wang}},
  \bibinfo {author} {\bibfnamefont {M.-H.}\ \bibnamefont {Whangbo}}, \ and\
  \bibinfo {author} {\bibfnamefont {X.~G.}\ \bibnamefont {Gong}},\ }\bibfield
  {title} {\enquote {\bibinfo {title} {{Unified model of ferroelectricity
  induced by spin order}},}\ }\href {\doibase 10.1103/PhysRevB.88.054404}
  {\bibfield  {journal} {\bibinfo  {journal} {Phys. Rev. B}\ }\textbf {\bibinfo
  {volume} {88}},\ \bibinfo {pages} {054404} (\bibinfo {year}
  {2013})}\BibitemShut {NoStop}%
\bibitem [{\citenamefont {Xiao}\ \emph {et~al.}(2009)\citenamefont {Xiao},
  \citenamefont {Shi}, \citenamefont {Clougherty},\ and\ \citenamefont
  {Niu}}]{Xiao2009}%
  \BibitemOpen
  \bibfield  {author} {\bibinfo {author} {\bibfnamefont {D.}~\bibnamefont
  {Xiao}}, \bibinfo {author} {\bibfnamefont {J.}~\bibnamefont {Shi}}, \bibinfo
  {author} {\bibfnamefont {D.~P.}\ \bibnamefont {Clougherty}}, \ and\ \bibinfo
  {author} {\bibfnamefont {Q.}~\bibnamefont {Niu}},\ }\bibfield  {title}
  {\enquote {\bibinfo {title} {Polarization and adiabatic pumping in
  inhomogeneous crystals},}\ }\href {\doibase 10.1103/PhysRevLett.102.087602}
  {\bibfield  {journal} {\bibinfo  {journal} {Phys. Rev. Lett.}\ }\textbf
  {\bibinfo {volume} {102}},\ \bibinfo {pages} {087602} (\bibinfo {year}
  {2009})}\BibitemShut {NoStop}%
\bibitem [{\citenamefont {Sundaram}\ and\ \citenamefont
  {Niu}(1999)}]{Sundaram1999}%
  \BibitemOpen
  \bibfield  {author} {\bibinfo {author} {\bibfnamefont {G.}~\bibnamefont
  {Sundaram}}\ and\ \bibinfo {author} {\bibfnamefont {Q.}~\bibnamefont {Niu}},\
  }\bibfield  {title} {\enquote {\bibinfo {title} {Wave-packet dynamics in
  slowly perturbed crystals: gradient corrections and berry-phase effects},}\
  }\href {\doibase 10.1103/PhysRevB.59.14915} {\bibfield  {journal} {\bibinfo
  {journal} {Phys. Rev. B}\ }\textbf {\bibinfo {volume} {59}},\ \bibinfo
  {pages} {14915--14925} (\bibinfo {year} {1999})}\BibitemShut {NoStop}%
\bibitem [{\citenamefont {Xiao}\ \emph {et~al.}(2010)\citenamefont {Xiao},
  \citenamefont {Chang},\ and\ \citenamefont {Niu}}]{Xiao2010}%
  \BibitemOpen
  \bibfield  {author} {\bibinfo {author} {\bibfnamefont {D.}~\bibnamefont
  {Xiao}}, \bibinfo {author} {\bibfnamefont {M.-C.}\ \bibnamefont {Chang}}, \
  and\ \bibinfo {author} {\bibfnamefont {Q.}~\bibnamefont {Niu}},\ }\bibfield
  {title} {\enquote {\bibinfo {title} {Berry phase effects on electronic
  properties},}\ }\href {\doibase 10.1103/RevModPhys.82.1959} {\bibfield
  {journal} {\bibinfo  {journal} {Rev. Mod. Phys.}\ }\textbf {\bibinfo {volume}
  {82}},\ \bibinfo {pages} {1959--2007} (\bibinfo {year} {2010})}\BibitemShut
  {NoStop}%
\bibitem [{\citenamefont {Xiao}\ \emph {et~al.}(2005)\citenamefont {Xiao},
  \citenamefont {Shi},\ and\ \citenamefont {Niu}}]{Xiao2005}%
  \BibitemOpen
  \bibfield  {author} {\bibinfo {author} {\bibfnamefont {D.}~\bibnamefont
  {Xiao}}, \bibinfo {author} {\bibfnamefont {J.}~\bibnamefont {Shi}}, \ and\
  \bibinfo {author} {\bibfnamefont {Q.}~\bibnamefont {Niu}},\ }\bibfield
  {title} {\enquote {\bibinfo {title} {Berry phase correction to electron
  density of states in solids},}\ }\href {\doibase
  10.1103/PhysRevLett.95.137204} {\bibfield  {journal} {\bibinfo  {journal}
  {Phys. Rev. Lett.}\ }\textbf {\bibinfo {volume} {95}},\ \bibinfo {pages}
  {137204} (\bibinfo {year} {2005})}\BibitemShut {NoStop}%
\bibitem [{\citenamefont {Gao}\ \emph {et~al.}(2015)\citenamefont {Gao},
  \citenamefont {Yang},\ and\ \citenamefont {Niu}}]{Gao2015}%
  \BibitemOpen
  \bibfield  {author} {\bibinfo {author} {\bibfnamefont {Y.}~\bibnamefont
  {Gao}}, \bibinfo {author} {\bibfnamefont {S.~A.}\ \bibnamefont {Yang}}, \
  and\ \bibinfo {author} {\bibfnamefont {Q.}~\bibnamefont {Niu}},\ }\bibfield
  {title} {\enquote {\bibinfo {title} {Geometrical effects in orbital magnetic
  susceptibility},}\ }\href {\doibase 10.1103/PhysRevB.91.214405} {\bibfield
  {journal} {\bibinfo  {journal} {Phys. Rev. B}\ }\textbf {\bibinfo {volume}
  {91}},\ \bibinfo {pages} {214405} (\bibinfo {year} {2015})}\BibitemShut
  {NoStop}%
\bibitem [{\citenamefont {Xiao}\ \emph {et~al.}(2006)\citenamefont {Xiao},
  \citenamefont {Yao}, \citenamefont {Fang},\ and\ \citenamefont
  {Niu}}]{Xiao2006}%
  \BibitemOpen
  \bibfield  {author} {\bibinfo {author} {\bibfnamefont {D.}~\bibnamefont
  {Xiao}}, \bibinfo {author} {\bibfnamefont {Y.}~\bibnamefont {Yao}}, \bibinfo
  {author} {\bibfnamefont {Z.}~\bibnamefont {Fang}}, \ and\ \bibinfo {author}
  {\bibfnamefont {Q.}~\bibnamefont {Niu}},\ }\bibfield  {title} {\enquote
  {\bibinfo {title} {Berry-phase effect in anomalous thermoelectric
  transport},}\ }\href {\doibase 10.1103/PhysRevLett.97.026603} {\bibfield
  {journal} {\bibinfo  {journal} {Phys. Rev. Lett.}\ }\textbf {\bibinfo
  {volume} {97}},\ \bibinfo {pages} {026603} (\bibinfo {year}
  {2006})}\BibitemShut {NoStop}%
\bibitem [{\citenamefont {Culcer}\ \emph {et~al.}(2004)\citenamefont {Culcer},
  \citenamefont {Sinova}, \citenamefont {Sinitsyn}, \citenamefont {Jungwirth},
  \citenamefont {MacDonald},\ and\ \citenamefont {Niu}}]{Culcer2004}%
  \BibitemOpen
  \bibfield  {author} {\bibinfo {author} {\bibfnamefont {D.}~\bibnamefont
  {Culcer}}, \bibinfo {author} {\bibfnamefont {J.}~\bibnamefont {Sinova}},
  \bibinfo {author} {\bibfnamefont {N.~A.}\ \bibnamefont {Sinitsyn}}, \bibinfo
  {author} {\bibfnamefont {T.}~\bibnamefont {Jungwirth}}, \bibinfo {author}
  {\bibfnamefont {A.~H.}\ \bibnamefont {MacDonald}}, \ and\ \bibinfo {author}
  {\bibfnamefont {Q.}~\bibnamefont {Niu}},\ }\bibfield  {title} {\enquote
  {\bibinfo {title} {Semiclassical spin transport in spin-orbit-coupled
  bands},}\ }\href {\doibase 10.1103/PhysRevLett.93.046602} {\bibfield
  {journal} {\bibinfo  {journal} {Phys. Rev. Lett.}\ }\textbf {\bibinfo
  {volume} {93}},\ \bibinfo {pages} {046602} (\bibinfo {year}
  {2004})}\BibitemShut {NoStop}%
\bibitem [{\citenamefont {Gao}\ and\ \citenamefont {Xiao}(2018)}]{Gao2018a}%
  \BibitemOpen
  \bibfield  {author} {\bibinfo {author} {\bibfnamefont {Y.}~\bibnamefont
  {Gao}}\ and\ \bibinfo {author} {\bibfnamefont {D.}~\bibnamefont {Xiao}},\
  }\bibfield  {title} {\enquote {\bibinfo {title} {Orbital magnetic quadrupole
  moment and nonlinear anomalous thermoelectric transport},}\ }\href {\doibase
  10.1103/PhysRevB.98.060402} {\bibfield  {journal} {\bibinfo  {journal} {Phys.
  Rev. B}\ }\textbf {\bibinfo {volume} {98}},\ \bibinfo {pages} {060402}
  (\bibinfo {year} {2018})}\BibitemShut {NoStop}%
\bibitem [{\citenamefont {Gao}\ and\ \citenamefont {Xiao}(2019)}]{Gao2019}%
  \BibitemOpen
  \bibfield  {author} {\bibinfo {author} {\bibfnamefont {Y.}~\bibnamefont
  {Gao}}\ and\ \bibinfo {author} {\bibfnamefont {D.}~\bibnamefont {Xiao}},\
  }\bibfield  {title} {\enquote {\bibinfo {title} {Nonreciprocal directional
  dichroism induced by the quantum metric dipole},}\ }\href {\doibase
  10.1103/PhysRevLett.122.227402} {\bibfield  {journal} {\bibinfo  {journal}
  {Phys. Rev. Lett.}\ }\textbf {\bibinfo {volume} {122}},\ \bibinfo {pages}
  {227402} (\bibinfo {year} {2019})}\BibitemShut {NoStop}%
\bibitem [{\citenamefont {Gao}\ \emph {et~al.}(2017)\citenamefont {Gao},
  \citenamefont {Yang},\ and\ \citenamefont {Niu}}]{Gao2017}%
  \BibitemOpen
  \bibfield  {author} {\bibinfo {author} {\bibfnamefont {Y.}~\bibnamefont
  {Gao}}, \bibinfo {author} {\bibfnamefont {S.~A.}\ \bibnamefont {Yang}}, \
  and\ \bibinfo {author} {\bibfnamefont {Q.}~\bibnamefont {Niu}},\ }\bibfield
  {title} {\enquote {\bibinfo {title} {Intrinsic relative magnetoconductivity
  of nonmagnetic metals},}\ }\href {\doibase 10.1103/PhysRevB.95.165135}
  {\bibfield  {journal} {\bibinfo  {journal} {Phys. Rev. B}\ }\textbf {\bibinfo
  {volume} {95}},\ \bibinfo {pages} {165135} (\bibinfo {year}
  {2017})}\BibitemShut {NoStop}%
\bibitem [{\citenamefont {Jian-Hui}\ \emph {et~al.}(2013)\citenamefont
  {Jian-Hui}, \citenamefont {Hua}, \citenamefont {Qian},\ and\ \citenamefont
  {Jun-Ren}}]{jian2013topological}%
  \BibitemOpen
  \bibfield  {author} {\bibinfo {author} {\bibfnamefont {Z.}~\bibnamefont
  {Jian-Hui}}, \bibinfo {author} {\bibfnamefont {J.}~\bibnamefont {Hua}},
  \bibinfo {author} {\bibfnamefont {N.}~\bibnamefont {Qian}}, \ and\ \bibinfo
  {author} {\bibfnamefont {S.}~\bibnamefont {Jun-Ren}},\ }\bibfield  {title}
  {\enquote {\bibinfo {title} {Topological invariants of metals and the related
  physical effects},}\ }\href@noop {} {\bibfield  {journal} {\bibinfo
  {journal} {Chinese Physics Letters}\ }\textbf {\bibinfo {volume} {30}},\
  \bibinfo {pages} {027101} (\bibinfo {year} {2013})}\BibitemShut {NoStop}%
\bibitem [{\citenamefont {Lapa}\ and\ \citenamefont {Hughes}(2019)}]{Lapa2019}%
  \BibitemOpen
  \bibfield  {author} {\bibinfo {author} {\bibfnamefont {M.~F.}\ \bibnamefont
  {Lapa}}\ and\ \bibinfo {author} {\bibfnamefont {T.~L.}\ \bibnamefont
  {Hughes}},\ }\bibfield  {title} {\enquote {\bibinfo {title} {Semiclassical
  wave packet dynamics in nonuniform electric fields},}\ }\href {\doibase
  10.1103/PhysRevB.99.121111} {\bibfield  {journal} {\bibinfo  {journal} {Phys.
  Rev. B}\ }\textbf {\bibinfo {volume} {99}},\ \bibinfo {pages} {121111}
  (\bibinfo {year} {2019})}\BibitemShut {NoStop}%
\bibitem [{\citenamefont {Chamon}\ \emph {et~al.}(2008)\citenamefont {Chamon},
  \citenamefont {Hou}, \citenamefont {Jackiw}, \citenamefont {Mudry},
  \citenamefont {Pi},\ and\ \citenamefont {Schnyder}}]{Chamon2008}%
  \BibitemOpen
  \bibfield  {author} {\bibinfo {author} {\bibfnamefont {C.}~\bibnamefont
  {Chamon}}, \bibinfo {author} {\bibfnamefont {C.-Y.}\ \bibnamefont {Hou}},
  \bibinfo {author} {\bibfnamefont {R.}~\bibnamefont {Jackiw}}, \bibinfo
  {author} {\bibfnamefont {C.}~\bibnamefont {Mudry}}, \bibinfo {author}
  {\bibfnamefont {S.-Y.}\ \bibnamefont {Pi}}, \ and\ \bibinfo {author}
  {\bibfnamefont {A.~P.}\ \bibnamefont {Schnyder}},\ }\bibfield  {title}
  {\enquote {\bibinfo {title} {Irrational versus rational charge and statistics
  in two-dimensional quantum systems},}\ }\href {\doibase
  10.1103/PhysRevLett.100.110405} {\bibfield  {journal} {\bibinfo  {journal}
  {Phys. Rev. Lett.}\ }\textbf {\bibinfo {volume} {100}},\ \bibinfo {pages}
  {110405} (\bibinfo {year} {2008})}\BibitemShut {NoStop}%
\bibitem [{\citenamefont {Seradjeh}\ \emph {et~al.}(2008)\citenamefont
  {Seradjeh}, \citenamefont {Weeks},\ and\ \citenamefont
  {Franz}}]{Seradjeh2008}%
  \BibitemOpen
  \bibfield  {author} {\bibinfo {author} {\bibfnamefont {B.}~\bibnamefont
  {Seradjeh}}, \bibinfo {author} {\bibfnamefont {C.}~\bibnamefont {Weeks}}, \
  and\ \bibinfo {author} {\bibfnamefont {M.}~\bibnamefont {Franz}},\ }\bibfield
   {title} {\enquote {\bibinfo {title} {Fractionalization in a square-lattice
  model with time-reversal symmetry},}\ }\href {\doibase
  10.1103/PhysRevB.77.033104} {\bibfield  {journal} {\bibinfo  {journal} {Phys.
  Rev. B}\ }\textbf {\bibinfo {volume} {77}},\ \bibinfo {pages} {033104}
  (\bibinfo {year} {2008})}\BibitemShut {NoStop}%
\bibitem [{\citenamefont {Benalcazar}\ \emph
  {et~al.}(2017{\natexlab{a}})\citenamefont {Benalcazar}, \citenamefont
  {Bernevig},\ and\ \citenamefont {Hughes}}]{Benalcazar2017a}%
  \BibitemOpen
  \bibfield  {author} {\bibinfo {author} {\bibfnamefont {W.~A.}\ \bibnamefont
  {Benalcazar}}, \bibinfo {author} {\bibfnamefont {B.~A.}\ \bibnamefont
  {Bernevig}}, \ and\ \bibinfo {author} {\bibfnamefont {T.~L.}\ \bibnamefont
  {Hughes}},\ }\bibfield  {title} {\enquote {\bibinfo {title} {Quantized
  electric multipole insulators},}\ }\href {\doibase 10.1126/science.aah6442}
  {\bibfield  {journal} {\bibinfo  {journal} {Science}\ }\textbf {\bibinfo
  {volume} {357}},\ \bibinfo {pages} {61--66} (\bibinfo {year}
  {2017}{\natexlab{a}})}\BibitemShut {NoStop}%
\bibitem [{\citenamefont {Benalcazar}\ \emph
  {et~al.}(2017{\natexlab{b}})\citenamefont {Benalcazar}, \citenamefont
  {Bernevig},\ and\ \citenamefont {Hughes}}]{Benalcazar2017}%
  \BibitemOpen
  \bibfield  {author} {\bibinfo {author} {\bibfnamefont {W.~A.}\ \bibnamefont
  {Benalcazar}}, \bibinfo {author} {\bibfnamefont {B.~A.}\ \bibnamefont
  {Bernevig}}, \ and\ \bibinfo {author} {\bibfnamefont {T.~L.}\ \bibnamefont
  {Hughes}},\ }\bibfield  {title} {\enquote {\bibinfo {title} {Electric
  multipole moments, topological multipole moment pumping, and chiral hinge
  states in crystalline insulators},}\ }\href {\doibase
  10.1103/PhysRevB.96.245115} {\bibfield  {journal} {\bibinfo  {journal} {Phys.
  Rev. B}\ }\textbf {\bibinfo {volume} {96}},\ \bibinfo {pages} {245115}
  (\bibinfo {year} {2017}{\natexlab{b}})}\BibitemShut {NoStop}%
\bibitem [{\citenamefont {Teo}\ and\ \citenamefont {Kane}(2010)}]{Teo2010}%
  \BibitemOpen
  \bibfield  {author} {\bibinfo {author} {\bibfnamefont {J.~C.~Y.}\
  \bibnamefont {Teo}}\ and\ \bibinfo {author} {\bibfnamefont {C.~L.}\
  \bibnamefont {Kane}},\ }\bibfield  {title} {\enquote {\bibinfo {title}
  {Topological defects and gapless modes in insulators and superconductors},}\
  }\href {\doibase 10.1103/PhysRevB.82.115120} {\bibfield  {journal} {\bibinfo
  {journal} {Phys. Rev. B}\ }\textbf {\bibinfo {volume} {82}},\ \bibinfo
  {pages} {115120} (\bibinfo {year} {2010})}\BibitemShut {NoStop}%
\bibitem [{\citenamefont {Lee}\ \emph {et~al.}(2020)\citenamefont {Lee},
  \citenamefont {Furusaki},\ and\ \citenamefont {Yang}}]{Lee2020}%
  \BibitemOpen
  \bibfield  {author} {\bibinfo {author} {\bibfnamefont {E.}~\bibnamefont
  {Lee}}, \bibinfo {author} {\bibfnamefont {A.}~\bibnamefont {Furusaki}}, \
  and\ \bibinfo {author} {\bibfnamefont {B.-J.}\ \bibnamefont {Yang}},\
  }\bibfield  {title} {\enquote {\bibinfo {title} {{Fractional charge bound to
  a vortex in two-dimensional topological crystalline insulators}},}\ }\href
  {\doibase 10.1103/PhysRevB.101.241109} {\bibfield  {journal} {\bibinfo
  {journal} {Phys. Rev. B}\ }\textbf {\bibinfo {volume} {101}},\ \bibinfo
  {pages} {241109} (\bibinfo {year} {2020})}\BibitemShut {NoStop}%
\end{thebibliography}
%

\end{document}